\documentclass{emulateapj}
\usepackage{multirow}
\usepackage{subfigure}
\usepackage{longtable}
\def\msun{$M_{\odot}$}

\def\xmm{{\it XMM-Newton}}

\shortauthors{Lin et al.}
\begin{document}

\title{Classification of X-ray Sources in the {\it XMM-Newton} Serendipitous Source Catalog: Objects of Special Interest}

\author{Dacheng Lin\altaffilmark{1,2,3}, Natalie A. Webb\altaffilmark{1,2}, Didier Barret\altaffilmark{1,2}}
\altaffiltext{1}{CNRS, IRAP, 9 avenue du Colonel Roche, BP 44346, F-31028 Toulouse Cedex 4, France}
\altaffiltext{2}{Universit\'{e} de Toulouse, UPS-OMP, IRAP, Toulouse, France}
\altaffiltext{3}{Department of Physics and Astronomy, University of Alabama, Box 870324, Tuscaloosa, AL 35487, USA, email: dlin@ua.edu}

\begin{abstract}

We analyze 18 sources that were found to show interesting properties of periodicity, very soft spectra, and/or large long-term variability in X-rays in our project of classification of sources from the 2XMMi-DR3 catalog but were poorly studied in the literature, in order to investigate their nature. Two hard sources show X-ray periodicities of $\sim$1.62 hr (\object{2XMM J165334.4-414423}) and $\sim$2.1 hr (\object{2XMM J133135.2-315541}) and are probably magnetic cataclysmic variables. One source \object{2XMM J123103.2+110648} is an active galactic nucleus (AGN) candidate showing very soft X-ray spectra ($kT\sim 0.1$ keV) and exhibiting an intermittent $\sim$3.8 hr quasi-periodic oscillation. There are six other very soft sources (with $kT < 0.2$ keV), which might be in other galaxies with luminosities between $\sim$$10^{38}$--$10^{42}$ erg s$^{-1}$. They probably represent a diverse group that might include objects such as ultrasoft AGNs and cool thermal disk emission from accreting intermediate-mass black holes. Six highly variable sources with harder spectra are probably in nearby galaxies with luminosities above $10^{37}$ erg s$^{-1}$ and thus are great candidates for extragalactic X-ray binaries. One of them (\object{2XMMi J004211.2+410429}, in M 31) is probably a new-born persistent source, having been X-ray bright and hard in 0.3--10 keV for at least four years since it was discovered to enter an outburst in 2007. Three highly variable hard sources appear at low galactic latitudes and have maximum luminosities below $\sim$$10^{34}$ erg s$^{-1}$ if they are in our Galaxy, thus great candidates for cataclysmic variables or very faint X-ray transients harboring a black hole or neutron star. Our interpretations of these sources can be tested with future long-term X-ray monitoring and multi-wavelength observations.

\end{abstract}


\section{INTRODUCTION}
\label{sec:intro}

\begin{figure*} 
\centering
\includegraphics[width=0.9\textwidth]{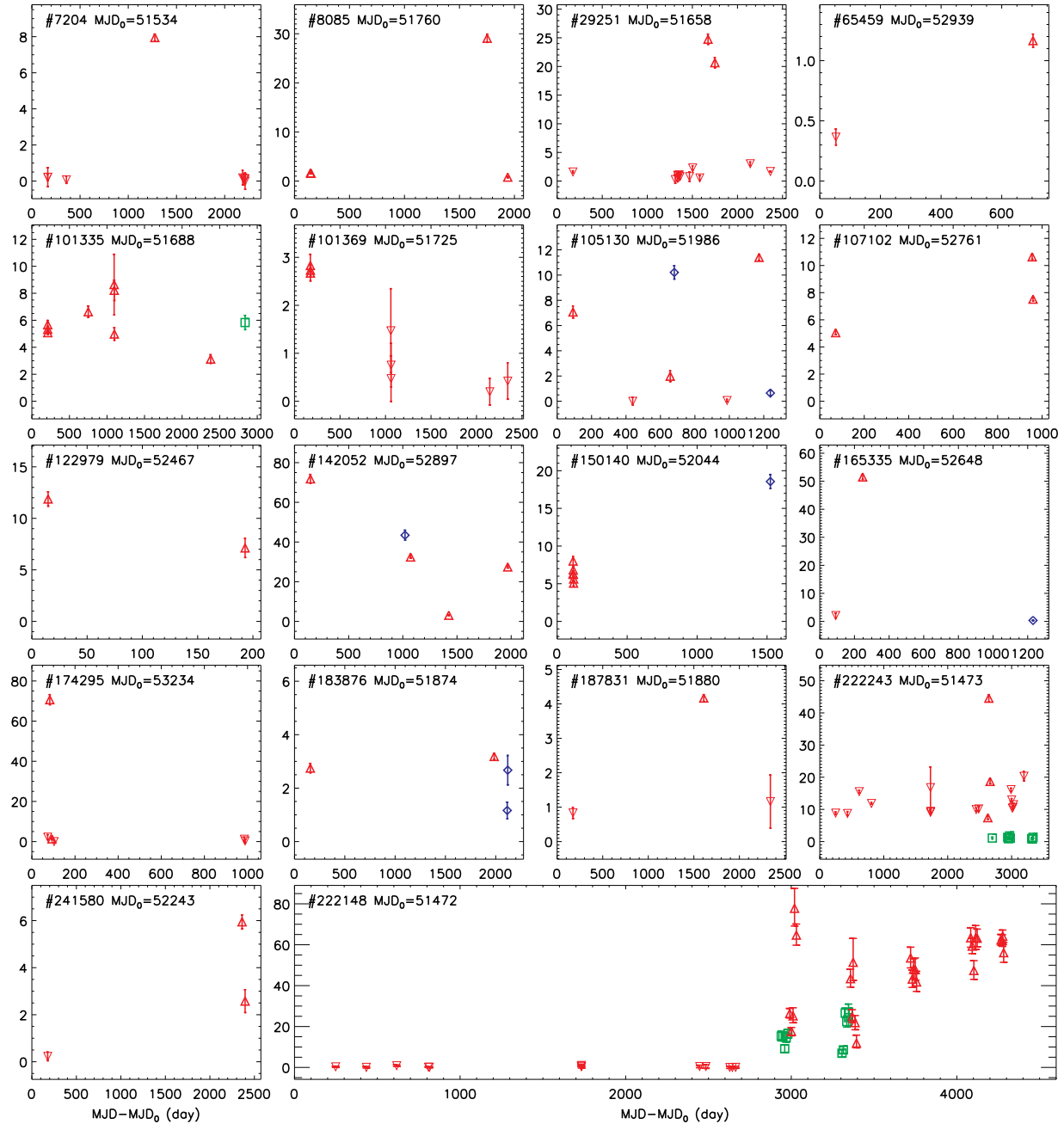}
\caption{The long-term flux curve (in units of $10^{-14}$ erg s$^{-1}$ cm$^{-2}$). The 0.2--4.5 keV fluxes are plotted for \xmm\ observations; those when the source was detected are from the 2XMMi-DR3 catalog (red triangles; for source \#222148, the 0.3--10 keV absorbed fluxes from our spectral fits are used (Table~\ref{tbl:spfit})), while those when the source was not detected were estimated from http://www.ledas.ac.uk/flix/flix.html (upside down triangles). For {\it Chandra} observations only detections from the CSC are plotted, with the 0.5--7 keV fluxes plotted for ACIS observations (blue diamonds) and 0.1--10 keV fluxes for High Resolution Camera (HRC) observations (green squares). \label{fig:srcltlc}}
\end{figure*}

In \citet[][LWB12 hereafter]{liweba2012}, we carried out source type
classifications of 4330 sources from the 2XMMi-DR3 catalog, which is
the largest X-ray source catalog ever produced, containing 353,191
X-ray source detections for 262,902 unique X-ray sources from 4953
pointed observations made by \xmm\ \citep{wascfy2009}. For about one
third of these 4330 sources we obtained reliable source types from the
literature. They mostly correspond to various types of stars, active
galactic nuclei (AGNs) and compact object (CO) systems containing
white dwarfs (WDs), neutron stars (NSs), and stellar-mass black holes
(BHs). These sources show different source properties in terms of the
X-ray spectral shape, X-ray variability, and the multi-wavelength
cross-correlation, based on which we came up with a method to classify
the rest of the sources into the above three main categories, i.e.,
stars, AGNs, and compact object systems. In brief, we first identified
star candidates as those with low X-ray-to-IR flux ratios and/or
frequent X-ray flares and then identified compact object system
candidates from the remaining sources as those with properties hardly
seen in AGNs, such as soft X-ray spectra and large long-term
variability. The compact object system candidates identified in this
way could inevitably include exotic objects with high science value,
such as very soft AGNs like \object{GSN 069} \citep{misaro2013} and
tidal disruption events (TDEs), in which stars approaching a
supermassive BH (SMBH) is tidally disrupted and subsequently accreted
\citep{lioz1979,re1988}.

About 200 strong compact object system candidates were found. Some of
them were poorly studied in the literature. Because these objects are
important targets for studies of dense matter and accretion physics in
extreme environment and they are also a treasure for discovery of
objects with high science value, we have planned a series of companion
papers on these poorly studied compact object system candidates in
order to shed new light on their nature. While the other companion
papers focus on single sources \citep[e.g.,][]{licagr2011,liweba2013},
here we present a relatively large sample (18 in total), which were
selected because they exhibited interesting properties of periodicity,
very soft spectra, and/or large long-term variability. Some of their
general properties from LWB12 are given in Table~\ref{tbl:dercat},
including the number of source detections in the 2XMMi-DR3 catalog,
the number of the \xmm\ observations, the 0.2--4.5 keV flux variation
($V_{\rm var14}$), etc. Throughout the paper, we generally use the
unique source index from the 2XMMi-DR3 catalog (column SRCID) to refer
to the sources (Table~\ref{tbl:dercat}).

In this paper, we concentrate primarily on \xmm\ X-ray
observations. We only present those that best demonstrate the source
properties based on considerations of data quality, sampling of
spectral states, etc. Some sources are also observed by other X-ray
observatories, especially the {\it Chandra} X-ray Observatory
(CXO). We mostly refer to the {\it Chandra} Source Catalog \citep[CSC,
  release 1.1,][]{evprgl2010} for their behavior as observed with {\it
  Chandra}, except for one observation, which was analyzed to show the
period of source \#150140. In Section~\ref{sec:reduction}, we describe
the analysis of the selected \xmm\ and {\it Chandra} observations of
the sources. In Section~\ref{sec:res}, we present the results and
discuss the possible nature of each source. The conclusions of our
study are given in Section~\ref{sec:conclusion}. Throughout the paper,
we assume a flat universe with the Hubble constant $H_0$=73 km
s$^{-1}$ Mpc$^{-1}$ and the matter density $\Omega_{\rm M}$=0.27.

\section{DATA ANALYSIS}
\label{sec:reduction}

\begin{figure*} 
\centering
\includegraphics[width=6.0in]{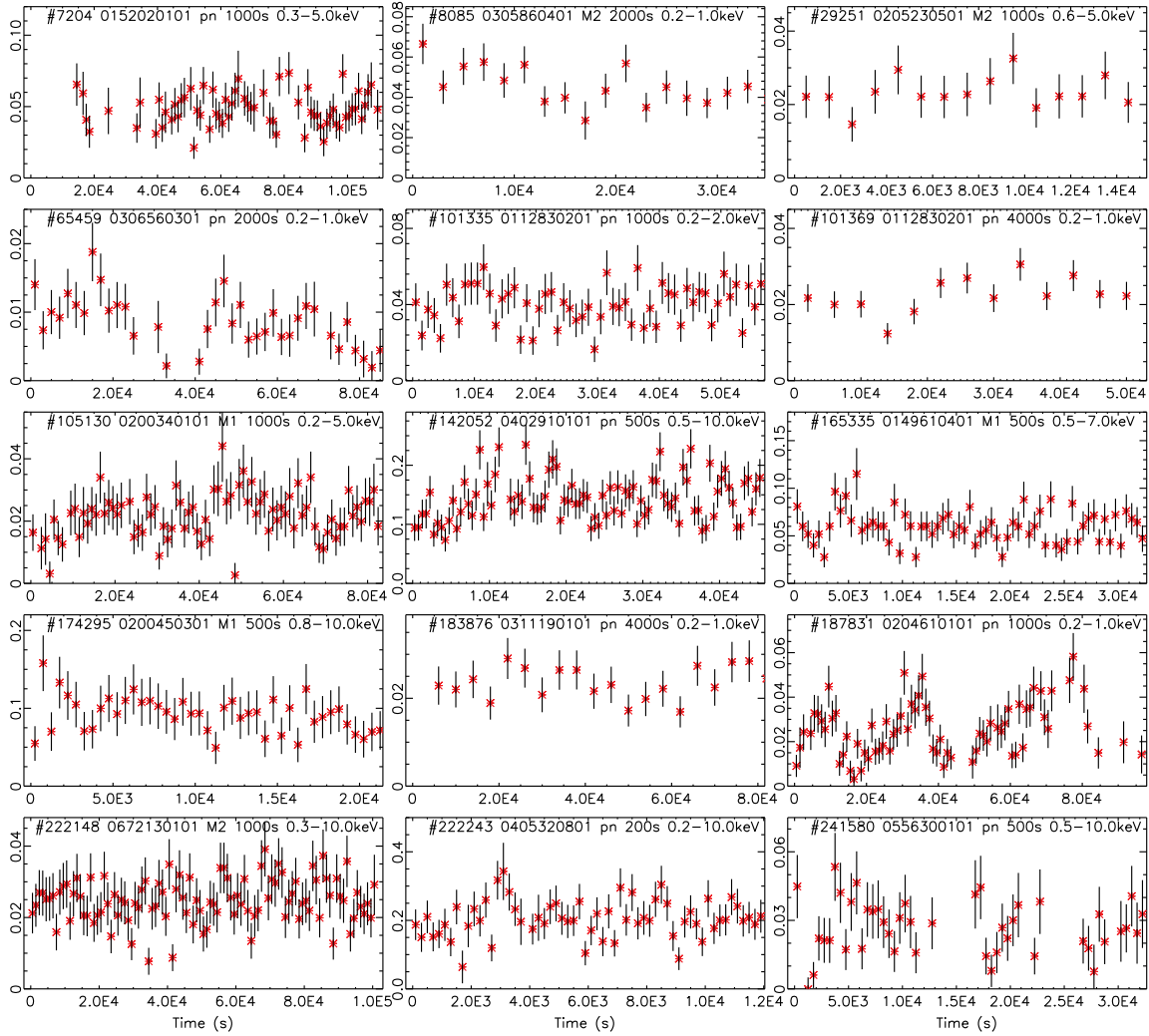}
\caption{The light curves (in units of cts s$^{-1}$) of bright detections. In each panel, we note the SRCID number, the observation ID, the EPIC camera, the bin time, and the energy band used to produce each light curve. \label{fig:srclc}}
\end{figure*}

\tabletypesize{\tiny}
\setlength{\tabcolsep}{0.01in}
\begin{deluxetable*}{rlcccccccccccccccccc}
\centering
\tablecaption{General Source Properties from LWB12 \label{tbl:dercat}}
\tablewidth{0pt}
\tablehead{SRCID & \colhead{2XMMi-DR3} & \colhead{ePos} &\colhead{GLAT}&\colhead{NObs} & \colhead{DObs} &\colhead{NDet} &\colhead{DDet} & \colhead{Type} & \colhead{SrcChar} &  \colhead{Dxo} &\colhead{$R2$} & \colhead{Rxo} & \colhead{Dxir} & \colhead{$K_{\rm s}$} & \colhead{Rxir} & \colhead{Fmax} & \colhead{Vvar14}   & \colhead{RC3Name} & \colhead{RC3SepR}\\
(1) &\colhead{(2)} &(3) &(4) &(5) &(6) &(7) &(8) &(9) &(10) &(11) &(12) &(13) &(14) &(15)&(16) &(17) &(18) &(19) & (20)
}
\startdata
7204 &  2XMM J004705.4-251942 &  0.4 & $-88.0$ & 9 &  2047.70 &  1 &     0.00 &    CO &   S,V,E &\nodata&\nodata& $ 0.71$ &\nodata&\nodata& $ 0.01$ & $8.8\pm0.5$ &    22.68 &          NGC253 & 0.74 \\
8085 &  2XMM J005412.9-373309 &  0.3 & $-79.6$ &  4 &  1794.51 &  4 &  1794.51 &    CO &   S,V,E &\nodata&\nodata& $ 1.26$ &\nodata&\nodata& $ 0.56$ & $31\pm2$&    36.33 &          NGC300 & 1.11 \\
29251 &  2XMM J031820.8-663035 &  0.3 & $-44.6$ & 14 &  2189.90 &  2 &    76.19 &    CO &   S,V,E &\nodata&\nodata& $ 1.26$ &\nodata&\nodata& $ 0.56$ & $31\pm2$ &    29.50 &         NGC1313 & 0.18 \\
65459 &  2XMM J080703.5-763248 &  0.5 & $-22.1$ & 2 &   650.17 &  1 &     0.00 &    CO &     S &\nodata&\nodata& $-0.12$ &\nodata&\nodata& $-0.82$  & $1.3\pm0.1$ &     3.19 &\nodata&\nodata\\
101335 &  2XMM J121028.9+391748 &  0.2 & $75.1$ &  8 &  2169.02 &  8 &  2169.02 &    CO &     S &\nodata&\nodata& $ 0.70$ &\nodata&\nodata& $-0.00$ & $8\pm2$ &     2.63 &\nodata&\nodata\\
101369 &  2XMM J121035.0+393123 &  0.3 & $75.0$ &  8 &  2169.00 &  3 &     0.78 &    CO &     S &  0.5 & 19.33 & $-0.40$ &\nodata&\nodata& $-0.43$ & $3\pm1$ &     5.09 &\nodata&\nodata\\
105130 &  2XMM J122628.8+333626 &  0.3 & $81.6$ &  5 &  1080.48 &  3 &  1080.48 &    CO &   S,V,E &\nodata&\nodata& $ 0.85$ &\nodata&\nodata& $ 0.15$ & $11.9\pm0.4$ &    64.92 &         NGC4395 & 1.65 \\
107102 &  2XMM J123103.2+110648 &  0.2 & $73.3$ &  3 &   887.86 &  3 &   887.86 &    CO &     S &  3.6 & 19.72 & $ 0.31$ &\nodata&\nodata& $ 0.12$ & $11.2\pm0.3$ &     2.11 &\nodata&\nodata \\
122979 &  2XMM J133135.2-315541 &  0.5 & $30.2$ &  2 &   178.26 &  2 &   178.26 &    CO &     H &\nodata&\nodata& $ 1.33$ &\nodata&\nodata& $ 0.63$ & $37\pm3$ &     1.67 &\nodata&\nodata\\
142052 &  2XMM J154951.7-541630 &  0.2 & $-0.0$ &  4 &  1821.84 &  4 &  1821.84 &    CO &    V,G &  0.7 & 16.82 & $ 0.33$ &  1.3 & 13.76 & $ 0.69$ & $172\pm8$&    23.22 &\nodata&\nodata\\
150140 &  2XMM J165334.4-414423 &  0.2 & $1.3$  &  6 &     4.51 &  6 &     4.51 &    CO &    H,G &\nodata&\nodata& $ 1.23$ &\nodata&\nodata& $ 0.53$ & $29\pm2$ &     1.58 &\nodata&\nodata\\
165335 &  2XMM J181920.1-204541 &  0.4 & $-2.6$ &  2 &   155.58 &  1 &     0.00 &    CO &    V,G &  0.6 & 20.24 & $ 1.45$ &  0.6 & 12.87 & $ 0.08$ & $97\pm3$ &    24.46 &\nodata&\nodata\\
174295 &  2XMM J203353.6+410717 &  0.4 & $0.6$  &  6 &   915.50 &  2 &     9.97 &    CO &    V,G &  0.7 & 19.70 & $ 1.60$ &  0.2 & 14.01 & $ 0.91$ & $225\pm13$ &    52.24 &\nodata&\nodata\\
183876 &  2XMM J223545.1-260451 &  0.3 & $-59.7$&  2 &  1827.13 &  2 &  1827.13 &   ULX &    S,E &\nodata&\nodata& $ 0.25$ &\nodata&\nodata& $-0.45$ & $3.0\pm0.3$ &     1.16 &         NGC7314 & 0.82 \\
187831 &  2XMM J231818.7-422237 &  0.4 & $-65.7$&  3 &  2166.30 &  1 &     0.00 &   ULX &    S,E &\nodata&\nodata& $ 0.40$ &\nodata&\nodata& $-0.30$ & $4.2\pm0.1$ &     5.01 &         NGC7582 & 0.92 \\
222148 & 2XMMi J004211.2+410429 &  0.3 & $-21.8$& 18 &  2782.71 &  5 &    39.63 &    CO &    V,E &\nodata&\nodata& $ 1.77$ &\nodata&\nodata& $ 1.07$ & $100\pm13$ &   152.20 &          NGC224 & 0.15 \\
222243 & 2XMMi J004246.0+411736 &  0.3 & $-21.5$& 19 &  2944.80 &  3 &    35.56 &   XGS &     E &\nodata&\nodata& $ 1.86$ &  0.8 & 14.43 & $ 0.81$ & $122\pm4$ &     6.00 &          NGC224 & 0.02 \\
241580 & 2XMMi J123047.0+413651 &  0.4 & $74.9$ &  3 &  2217.86 &  2 &    33.89 &    CO &    V,E &\nodata&\nodata& $ 0.80$ &\nodata&\nodata& $ 0.10$ & $11\pm1$ &    16.10 &         NGC4490 & 0.87
\enddata 
\tablecomments{Columns are as follows. (1): 2XMM-DR3 unique source index; (2): 2XMMi-DR3 Source designation; (3): X-ray $1$-$\sigma$ positional error for each coordinate, in units of arcsec; (4): Galactic latitude; (5)--(6): Number of \xmm\ observations and the range of days; (7)--(8): Number of detections from the 2XMMi-DR3 catalog and the range of days; (9): Source type (``CO'': candidate compact object system based on source properties given in column (10) (see LWB12); ``ULX'': candidate ultraluminous off-nuclear X-ray source; ``XGS'': candidate non-nuclear extragalactic source); (10) The source characteristics based on which we did not classify the source as an AGN (``S'': soft (HR1$<$$-0.4$, HR2$<$$-0.5$, HR3$<$$-0.7$, or HR4$<$$-0.8$, where HR1--HR4 are hardness ratios; see LWB12); ``H'': hard ($-0.1$$<$HR3$<$0.5 and $-0.25$$<$HR4$<$0.1); ``V'': highly variable $V_{\rm var14}$$>$10; ``E'': non-nuclear extra-galactic;  ``G'': in Galactic plane ($|b|$$<$10)); (11) X-ray-optical separation (arcsec); (12): $R2$-band magnitude; (13) X-ray-to-optical flux ratio logarithm $\log(F_{\rm X}/F_{\rm O})$, where $F_{\rm X}$ is the maximum 0.2-12 keV flux and $F_{\rm O}$ is the optical flux defined as $\log(F_{\rm O})=-R2/2.5-5.37$ (when no optical counterpart, we assumed $R2=21$; see LWB12); (14): X-ray-IR separation (arcsec) (15): $K_{\rm s}$-band magnitude (16): X-ray-to-IR flux ratio logarithm $\log(F_{\rm X}/F_{\rm IR})$, where $F_{\rm IR}$ is the IR flux defined as $\log(F_{\rm IR})=-K_{\rm s}/2.5-6.95$ (when no IR counterpart, we assumed $K_{\rm s}=15.3$, the 3-$\sigma$ limiting sensitivity of the $K_{\rm s}$ band in the 2MASS; see LWB12); (17): The maximum 0.2--12 keV flux, in units of 10$^{-14}$ erg s$^{-1}$ cm$^{-2}$; (18): Flux variation factor in the 0.2--4.5 keV band; (19) The source name of the RC3 match; (20) The ratio of the angular separation to the $D_{25}$ isophote elliptical radius.}
\end{deluxetable*}

\tabletypesize{\scriptsize}
\setlength{\tabcolsep}{0.01in}
\begin{deluxetable}{rccccccccc}
\tablecaption{The Log of the \xmm\ and {\it Chandra} Observations Analyzed\label{tbl:obslog}}
\tablewidth{0pt}
\tablehead{SRCID & \colhead{Obs. ID} &\colhead{Date}&\colhead{off-axis angle } &\colhead{Exposure} &\colhead{$r_{\rm src}$}   \\
 & & & (arcmin) & (ks)& (arcsec)\\
               &  &             &\colhead{pn/M1/M2}&\colhead{pn/M1/M2} 
}
\startdata
  7204 & 0152020101 & 2003-06-20 & 8.1/7.3/6.6 & 58.4/77.7/78.2 & 15 \\
\hline
\multirow{2}{*}{8085} & 0305860401 & 2005-05-22 & -/-/11.6 & -/-/34.7 & 25\\
     & 0112800201 & 2000-12-26 & 11.5/11.6/12.6 & 27.1/33.4/33.6 & 15\\
\hline
29251 & 0205230501 & 2004-11-23 & 6.4/5.4/5.3 & 12.5/15.5/15.5 & 12 \\
\hline
65459 & 0306560301 & 2005-09-30 & 2.7/1.7/1.8 & 63.1/86.9/86.6 & 12\\
\hline
\multirow{2}{*}{101335} & 0112830201 & 2000-12-22 & 6.8/6.7/7.8 & 50.8/58.9/58.9 & 13 \\
       & 0402660201 & 2006-11-29 & -/6.7/7.8 &  -/29.3/28.8 & 10\\
\hline
101369 & 0112830201 & 2000-12-22 & 7.0/6.9/5.9 & 50.9/58.9/58.9 & 14\\
\hline
105130 & 0200340101 & 2004-06-03 & 6.9/7.9/8.0 &  60.4/75.7/76.7 & 20\\
\hline
\multirow{3}{*}{107102} & 0145800101 & 2003-07-13 & 7.5/7.1/8.1 & 45.4/58.0/61.4 & 25\\
                        & 0306630101 & 2005-12-13 & 6.8/-/5.9   & 54.8/-/68.7    & 25\\
                        & 0306630201 & 2005-12-17 & 6.8/-/5.9   & 80.8/-/92.2    & 25\\
\hline
\multirow{2}{*}{122979} & 0105261401 & 2002-07-27 & 8.0/8.3/- & 10.2/17.4/- & 15 \\
                        & 0105261701 & 2003-01-21 & 17.4/-/- & 15.3/-/- & 25 \\
\hline
\multirow{3}{*}{142052} &0203910101 & 2004-02-08 & 10.1/9.1/8.6 & 5.9/8.7/8.7 & 25 \\
& 0402910101 & 2006-08-21 & 8.1/9.1/9.5 & 38.8/45.6/45.5 & 25 \\
& 0410581901 & 2007-08-09 & 8.4/9.3/9.9 & 13.2/16.3/16.3 & 13\\
& 0560181101 & 2009-02-04 & 10.2/9.3/8.8 & 48.5/57.1/57.2 & 25 \\
\hline
\multirow{7}{*}{150140} & 0109490101 & 2001-09-05 & 7.8/8.4/9.3 & 27.4/32.9/32.9 & 15 \\
                        & 0109490201 & 2001-09-06 & 7.8/8.4/9.3 & 15.9/20.1/20.4 & 15 \\
                        & 0109490301 & 2001-09-07 & 7.8/8.4/9.3 & 28.4/34.0/34.1 & 15 \\
                        & 0109490401 & 2001-09-08 & 7.8/8.5/9.3 & 17.9/30.5/30.7 & 15 \\
                        & 0109490501 & 2001-09-09 & 7.8/8.4/9.3 & 25.4/30.7/30.7 & 15 \\
                        & 0109490601 & 2001-09-10 & 7.8/8.4/9.3 & 27.1/32.5/32.6 & 15 \\
                        & 6291(CXO)  & 2005-07-16 & 9.2  & 44 & 10 \\
\hline
165335 & 0149610401 & 2003-09-14 & 5.6/4.7/5.4 & 27.3/32.0/32.0 & 13\\
\hline
174295 & 0200450301 &2004-11-09 & 13.4/13.5/12.4 & -/21.4/21.4 & 25 \\
\hline
183876 & 0311190101 & 2006-05-04 & 8.2/8.2/9.2 & 67.7/82.6/82.7 & 12\\
\hline
187831 & 0204610101 & 2005-04-30 & 2.0/1.0/1.0 & 63.0/77.7/78.0 & 12\\
\hline
\multirow{24}{*}{222148}& 0505720201 & 2007-12-29 & -/13.1/12.7 & -/26.8/26.9 & 20 \\
& 0505720301 & 2008-01-08 & -/13.1/12.9 & -/26.5/26.5 & 20 \\
& 0505720401\tablenotemark{a} & 2008-01-18 & -/13.1/13.0 & -/21.6/21.7 & 20 \\
& 0505720501 & 2008-01-27 & -/13.1/- & -/20.0/- & 20 \\
& 0505720601 & 2008-02-07 & -/13.1/13.2 & -/21.2/21.2 & 20 \\
& 0551690201 & 2008-12-30 & -/13.2/12.8 & -/21.2/21.2 & 20 \\
& 0551690301 & 2009-01-09 & -/13.1/12.9 & -/21.1/21.1 & 20 \\
& 0551690401\tablenotemark{a} & 2009-01-15 & -/13.2/13.0 & -/6.1/5.8 & 20 \\
& 0551690501 & 2009-01-27 & -/13.2/- & -/20.6/- & 20 \\
& 0551690601\tablenotemark{a} & 2009-02-04 & -/13.1/13.1 & -/10.9/10.6 & 20 \\
& 0600660201 & 2009-12-28 & -/13.1/12.7 & -/18.1/18.2 & 20 \\
& 0600660301 & 2010-01-07 & -/13.1/12.9 & -/16.7/16.8 & 20 \\
& 0600660401 & 2010-01-15 & -/13.1/13.0 & -/16.7/16.7 & 20 \\
& 0600660501 & 2010-01-25 & -/13.1/- & -/19.1/- & 20 \\
& 0600660601\tablenotemark{a} & 2010-02-02 & -/13.1/13.1 & -/16.8/16.8 & 20 \\
& 0650560201 & 2010-12-26 & -/13.1/12.7 & -/24.1/24.3 & 20 \\
& 0650560301 & 2011-01-04 & -/13.1/12.9 & -/32.4/32.7 & 20 \\
& 0650560401 & 2011-01-14 & -/13.2/13.0 & -/18.8/19.4 & 20 \\
& 0650560501 & 2011-01-25 & -/13.2/- & -/23.3/- & 20 \\
& 0650560601 & 2011-02-03 & -/13.1/13.2 & -/23.3/23.2 & 20 \\
& 0672130101 & 2011-06-27 & -/-/12.8 & -/-/99.3 & 20 \\
& 0672130601 & 2011-07-05 & -/-/12.8 & -/-/91.5 & 20 \\
& 0672130701 & 2011-07-07 & -/-/12.8 & -/-/89.9 & 20 \\
& 0672130501\tablenotemark{b} & 2011-07-13 & -/-/12.9 & -/-/42.2 & 20 \\
\hline
222243 & 0405320801 & 2007-01-16 &2.6/1.7/2.4 & 9.9/13.4/13.4 &13\\
\hline
241580 & 0556300101 & 2008-05-19 & 3.8/2.8/2.7 & 18.7/30.0/30.6 & 13
\enddata 
\tablecomments{For the {\it Chandra} observation 6291 (marked by ``CXO''), the off-axis angle and the exposure refer to the ACIS detector.}
\tablenotetext{a}{In these observations, the source is right at the edge of the field of view of the MOS2 camera.}
\tablenotetext{b}{In this observation, the source is split by a CCD gap in the MOS2 camera}
\end{deluxetable}

\begin{figure*} 
\centering
\includegraphics[width=6.4in]{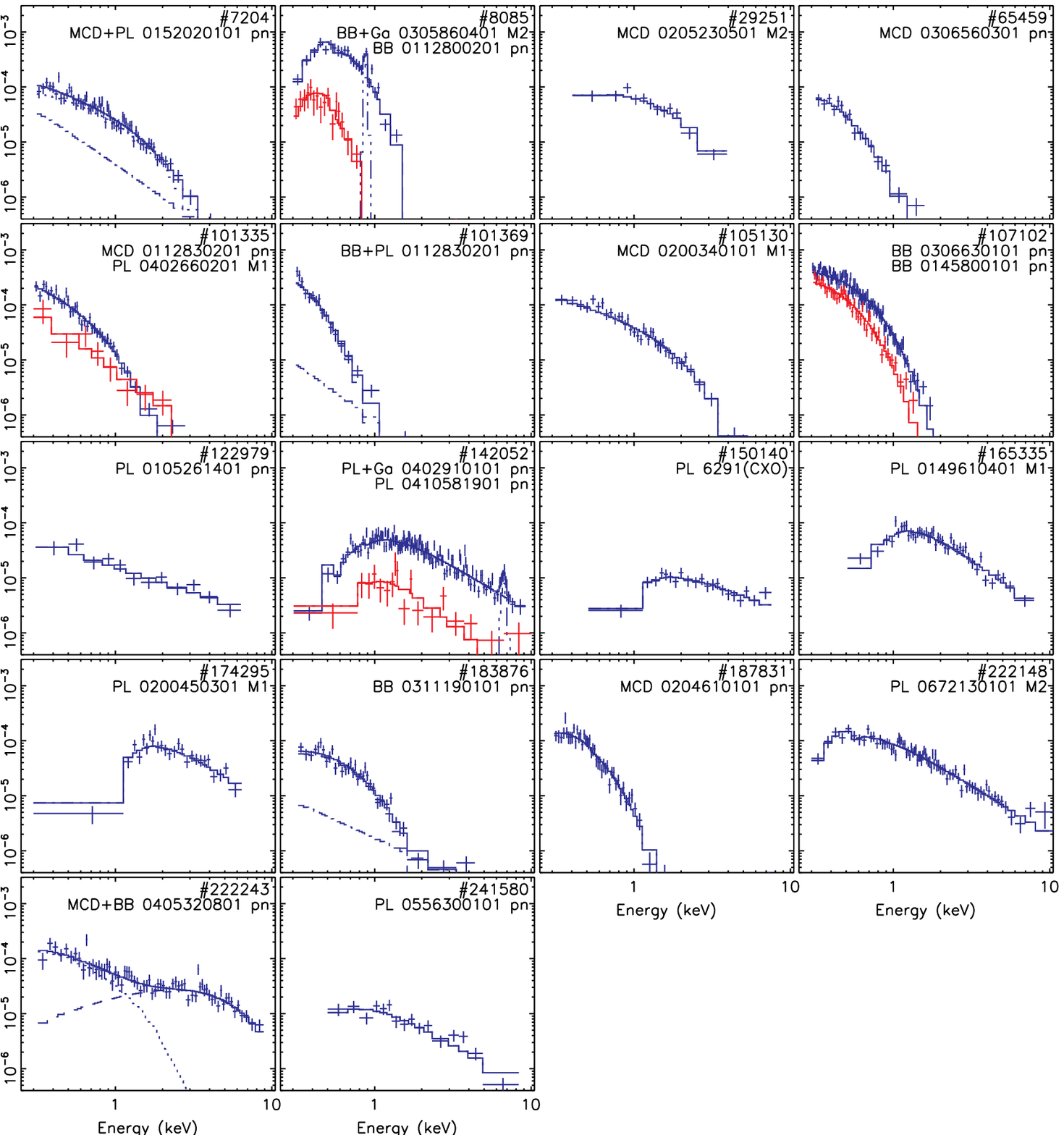}
\caption{The unfolded spectra (in units of photons cm$^{-2}$ s$^{-1}$ keV$^{-1}$) of some spectral fits. In each panel, we note the SRCID number, the spectral model, the observation ID, and the EPIC camera used in the plot. For panels with multiple spectra, the observation IDs are given in the order of the spectra from the top to the bottom. The dotted, dashed, dot-dashed,dot-dot-dot-dashed, and solid lines correspond to the MCD, BB, PL, Gaussian line, and the total models. \label{fig:srcspec}}
\end{figure*}

\begin{figure*} 
\centering
\includegraphics[width=5.8in]{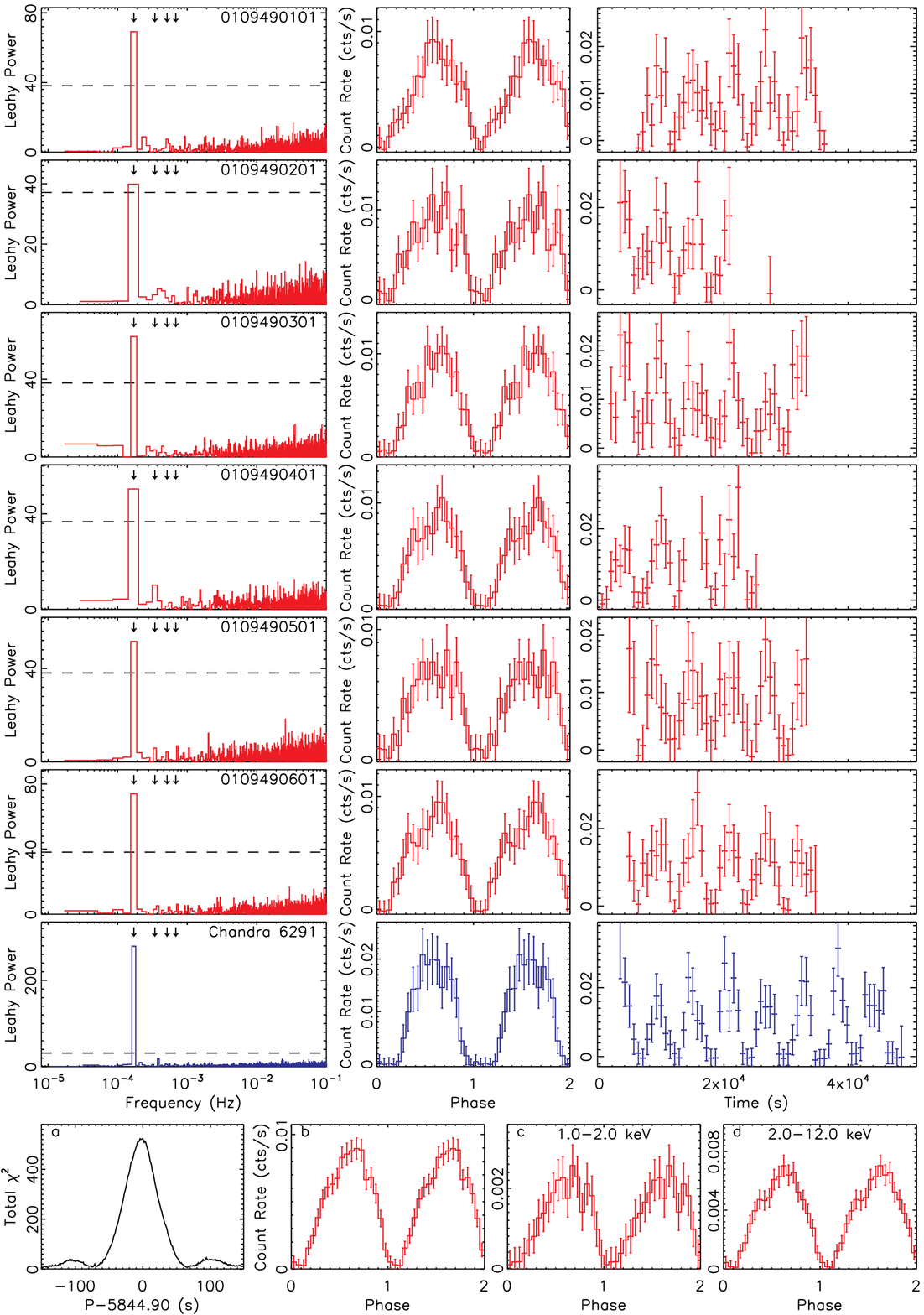}
\caption{The timing properties of source \#150140. The panels in each row (except the botton one) show the Leahy power (left), the light curve folded at a period of $P_0$=5844.90 s (middle), and the (unfolded) light curve (right) for each observation (1.0--12.0 keV). The arrows in the power plots mark the period of $P_0$ and harmonics ($P_0/2$, $P_0/3$, and $P_0/4$). The dashed lines indicate the 99.9\% confidence detection level  assuming blind search over all frequencies and Poisson noise as the underlying continuum. Only powers below 0.1 Hz are shown, and we see no powers above the 99.9\% confidence level above 0.1 Hz. The unfolded light curves are shifted in time to be aligned in phase. Panel (a) in the bottom row shows the total $\chi^2$ values from the fits with a constant to the 1.0--12.0 keV light curves folded at various tentative periods using all six \xmm\ observations 0109490101--601 (that is, the Chandra observation 6291 was not used). The 1.0--12.0 keV, 1.0--2.0 keV and 2.0--12.0 keV light curves folded at $P_0$ using also all six \xmm\ observations are given in panels (b)--(d), respectively. \label{fig:150140foldcurve}}
\end{figure*}

The details of the \xmm\ observations that we analyzed are given in
Table~\ref{tbl:obslog}. We used SAS 11.0.0 and the calibration files
of 2011 June for reprocessing the X-ray event files and follow-up
analysis. The data in strong background flare intervals are excluded
following the SAS thread for filtering against high background. We
extracted the source spectrum for all available cameras from a
circular region centered on the source, with the radius $r_{\rm src}$
for each observation given in Table~\ref{tbl:obslog}. The background
spectrum was extracted from a large circular region, typically with a
radius of 50$\arcsec$--100$\arcsec$, near the source in each
camera. The event selection criteria followed the default values in
the pipeline \citep[see Table~5 in][]{wascfy2009}. We rebinned the
spectra to have at least 20 counts in each bin so as to adopt the
$\chi^2$ statistic for the spectral fits, but for some very faint
detections, the spectra were not rebinned, and the fits were carried
out using the C statistic. Because most of our data have low
statistics, we fitted the spectra only with simple models: a powerlaw
(PL, {\it powerlaw} in XSPEC \citep{ar1996}), a multicolor disk (MCD,
{\it diskbb} in XSPEC), a single temperature blackbody (BB, {\it
  bbodyrad} in XSPEC), or their combinations. Sometimes, a Gaussian
emission line (described by the {\it gaussian} model in XSPEC) is
added when needed. All models included an interstellar medium
absorption, described by the {\it wabs} model in XSPEC.

For sources that we found to show periodic modulations in the light
curves from visual inspection, we created the Leahy power
\citep{ledael1983} for each observation to confirm the presence of
periodicity. We used pn light curves extracted from the source region
and binned at the frame readout time (MOS1 was used for the
observation 0105261401 of source \#122979 due to strong background
flares in pn).

To constrain the values of their periods and examine the profiles of
the modulation, we employed the epoch folding search technique. The
folded background-subtracted light curve for each camera is obtained
by subtracting the folded light curve from the source region by the
folded light curve from the background region (after being rescaled to
the size of the source region). The total folded light curve for
several cameras/observations are obtained by summing the folded
background-subtracted light curves for individual cameras weighted by
their total exposures (summed over all phase bins). All light curves
were barycenter-corrected.

We also analyzed one {\it Chandra} observation of source \#150140
(Table~\ref{tbl:obslog}) to check the persistence of its period. It
was carried out using the imaging array of the AXAF CCD Imaging
Spectrometer \citep[ACIS; ][]{bapiba1998}. We used the CIAO 4.4
package. The data were reprocessed to apply the latest calibration
(CALDB 4.4.8). We used radii of 10$\arcsec$ (Table~\ref{tbl:obslog})
and 50$\arcsec$ for the circular source and background regions,
respectively.

\section{RESULTS AND DISCUSSION}
\label{sec:res}
The long-term flux curves, light curves of bright detections, and
sample spectral fits of our sources are shown in
Figures~\ref{fig:srcltlc}--\ref{fig:srcspec}, respectively. The
spectral fit results are given in Table~\ref{tbl:spfit}.

\subsection{Periodic Hard Sources}
\begin{figure*} 
\centering
\includegraphics[width=5.8in]{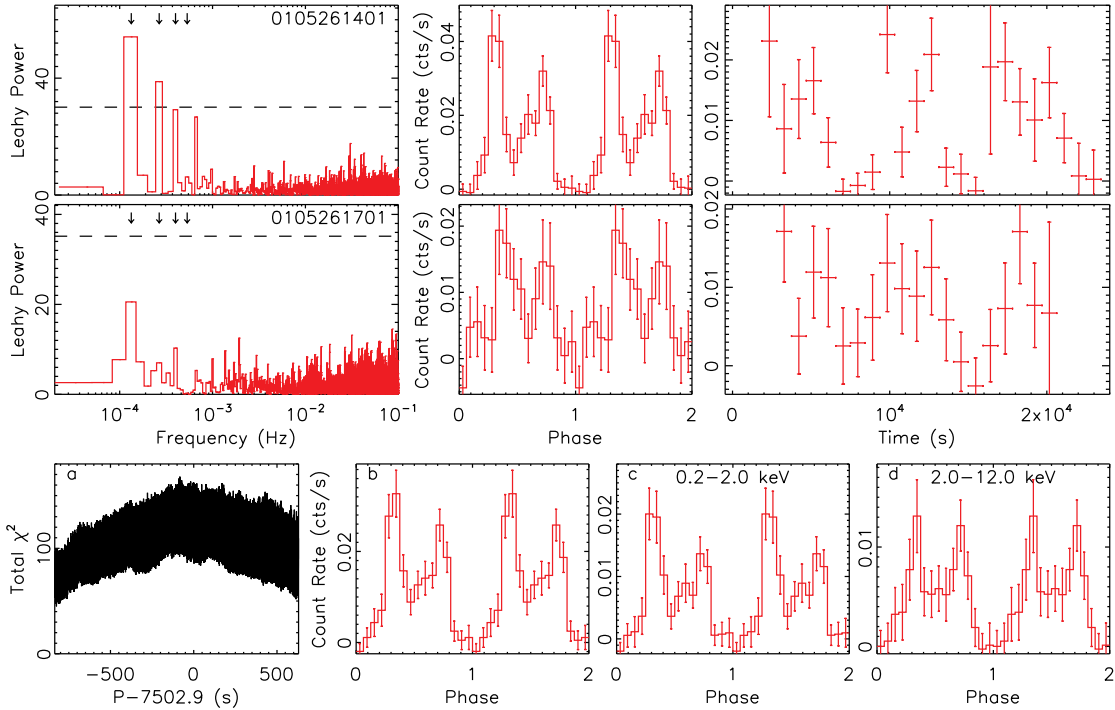}
\caption{The timing properties of source \#122979, similar to Figure~\ref{fig:150140foldcurve}. The energy band 0.2--12.0 keV is used for all plots except panels (c)--(d). Both observations are combined to search for the periodicity (panel (a)) and plot the folded light curves in panels (b)--(d). The Leahy power and the (unfolded) light curve for observation 0105261401 are from MOS1. \label{fig:122979foldcurve}}
\end{figure*}

We discover signs of coherent pulsations from three hard X-ray
sources, \#254026, \#150140 and \#122979. Source \#254026 was
presented in \citet{liweba2013}, and in the following we focus on the
latter two sources. Their X-ray spectra can be fitted with an absorbed
PL, with photon indices $\Gamma_{\rm PL}\sim 1$
(Table~\ref{tbl:spfit}).  Neither of them have optical or IR
counterparts found in the USNO-B1.0 Catalog \citep{moleca2003} or the
2MASS Point Source Catalog \citep[2MASS PSC,][]{cuskva2003} within
4$\arcsec$.

Source \#150140 appears on the outskirts of the young open cluster NGC
6231 \citep[about 1.64 kpc away and a few Myr
  old,][]{sagora2006}. There are six \xmm\ observations spanning only
five days in 2001 September, with $V_{\rm var14}$=1.6. The 0.3--10 keV
absorbed luminosity is about 7$\times$$10^{31}$ erg s$^{-1}$ if the
source is in this cluster or about 2$\times$$10^{33}$ erg s$^{-1}$ if
a source distance of 8.5 kpc is assumed. There is a {\it Chandra}
observation taken nearly four years later, with a similar spectrum
(Table~\ref{tbl:spfit}). The spectral fits indicate fairly strong
absorption ($N_{\rm H}\sim 0.6\times10^{22}$ cm$^{-1}$) toward the
source. Figure~\ref{fig:150140foldcurve}, which plots the timing
properties of the source, shows that there are X-ray modulations at a
period around 5845 s (i.e., 1.62 hr) in all the \xmm\ and {\it
  Chandra} observations. In the search for the period using the epoch
folding technique, we used only the \xmm\ observations, as their time
gap with {\it Chandra} observation is too large and adding the {\it
  Chandra} observation only results in many similar local maxima of
the total $\chi^2$ near the period. The pulse profile shows only a
single peak. The low phase has count rates consistent with zero but
seems to last longer in the {\it Chandra} observation (for about 25\%
of the period) than in the \xmm\ observations (for about 10\% of the
period). There is no significant variation of the pulse profile with
energy. Considering its low X-ray luminosity and hard X-ray spectra,
this source is probably a magnetic cataclysmic variable (CV), i.e., a
polar or a intermediate polar. With the period and the pulse profile
resembling those of some short-period one-pole polars \citep[e.g.,
  V347 Pav and GG Leo,][]{racrma2004}, source \#150140 is most likely
one such system. If it is really a CV, it is probably not in NGC 6231,
which seems too young to form a CV. This is also supported by the
larger column density inferred from our spectral fitting than that of
NGC 6231 \citep[(2--3)$\times10^{21}$ cm$^{-2}$,][]{mablde2004}.

Source \#122979 has a Galactic latitude of 30.2$\degr$ and thus is
probably nearby if it is in our Galaxy. It has two \xmm\ observations
separated by about half a year, with $V_{\rm var14}$=1.7. The second
observation has a very large off-axis angle of 17.4$\arcmin$, and the
source is only in the FOV of pn. The maximum 0.3--10 keV absorbed
luminosity, from the first observation, is about 3$\times$$10^{31}$
erg s$^{-1}$, assuming a source distance of 1 kpc. We found a possible
$\sim$2.1 hr periodicity (Figure~\ref{fig:122979foldcurve}), which,
however, needs to be confirmed with longer observations. The source
might be a magnetic CV, considering its probably low luminosity and
hard X-ray spectra.

\subsection{Very Soft Sources}
\begin{figure*} 
\centering
\includegraphics[width=5.8in]{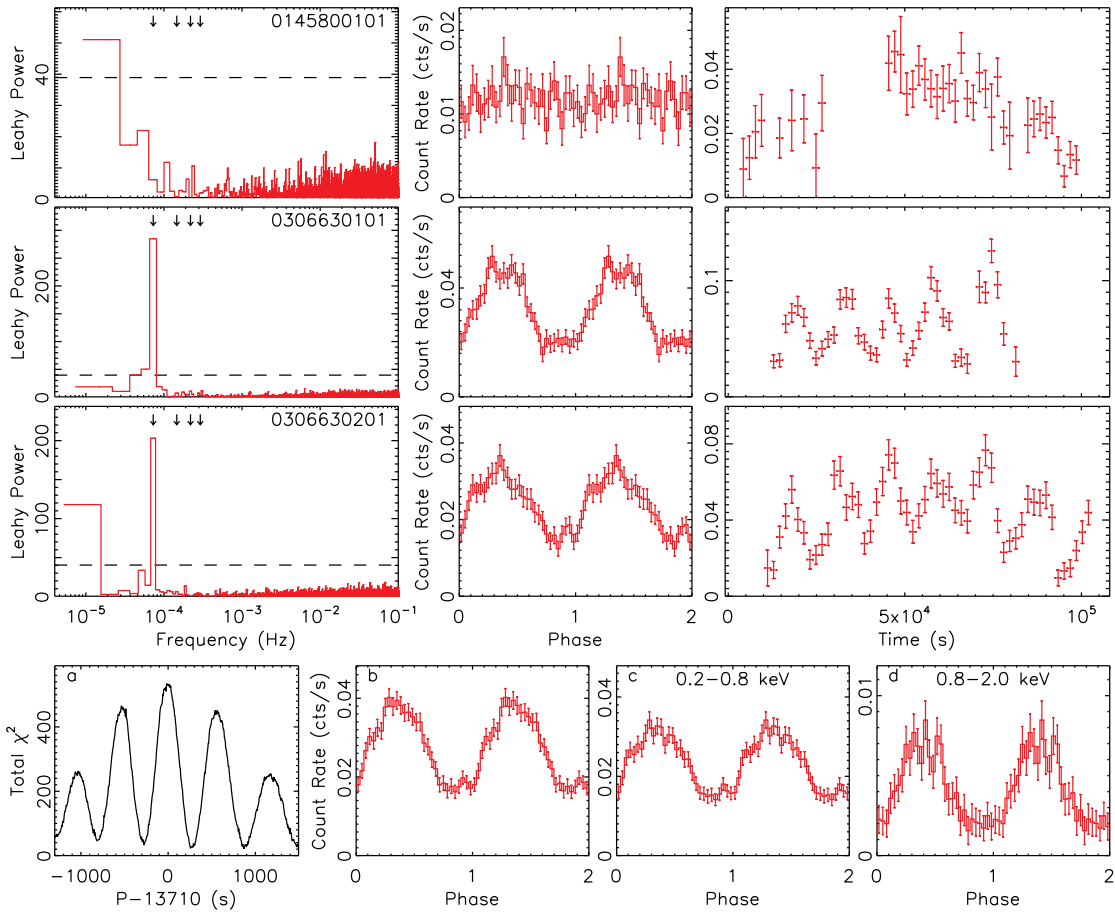}
\caption{The timing properties of source \#107102, similar to Figure~\ref{fig:150140foldcurve}. The energy band 0.2--2.0 keV is used for all plots except panels (c) and (d). Only the two observations 0306630101--201 are combined to search for the periodicity (panel (a)) and plot the folded light curves in panels (b)--(d). The power of observation 0145800101 was calculated using only the second segment of data (after 40 ks in the light curve plot in the top right panel) because earlier data suffer from frequent strong background flares.  \label{fig:107102foldcurve}}
\end{figure*}

We also found a handful of poorly studied very soft sources. Such
sources have been observed in our Milky Way, the Magellanic Clouds,
and nearby galaxies \citep[for a recent review,
  see][]{dikopr2010}. They are often referred to as supersoft (BB
temperature $kT_{\rm BB}\lesssim 0.1$ keV) or quasisoft ($kT_{\rm
  BB}\lesssim 0.3$ keV) X-ray sources. They most probably represent a
diverse group. If they are coincident with galactic nuclei, they could
be ultrasoft AGNs or TDEs \citep[e.g.,][]{licagr2011,misaro2013}. If
they are off-nuclear and have luminosities of $10^{35}$--$10^{38}$ erg
s$^{-1}$, they could be nuclear burning of material accreted by a WD
\citep{gr2000}, supernova remnants \citep[e.g.,][]{kosjwi2003},
central stars in planetary nebulae \citep[e.g.,][]{mekr2010}, or hot
stellar core remnants from tidal stripping of giants by SMBHs
\citep{digrmu2001}. Those with luminosities below $10^{34}$ erg
s$^{-1}$ could be thermally cooling isolated NSs
\citep[e.g.,][]{kavaan2002}, NS X-ray binaries in quiescence
\citep[e.g.,][]{weba2007}, or WDs (e.g., \object{WD 1254+223}, which
could be a binary \citep{bichgr2010}). Very soft sources with
luminosities above the Eddington limit of WDs
($\sim$2$\times$$10^{38}$ erg s$^{-1}$), such as ultraluminous
off-nuclear X-ray sources (ULXs, $>$$10^{39}$ erg s$^{-1}$), have two
popular explanations: cool thermal disk emission from accreting
intermediate-mass BHs \citep[IMBHs, $\sim$$10^2$--$10^5$
  \msun,][]{mifami2004,kodi2005,li2011} and optically thick outflows
from supercritical accretion viewed at high inclinations
\citep{mupesn2003,kipo2003,polifa2007,feso2011}. The spectral behavior
in these two scenarios is expected to be different: the former should
have the disk luminosity following the $L\propto T^4$ relation, and
the latter should have the observed outflow luminosity decreasing
slightly with increasing temperature \citep{polifa2007}. However,
neither trend has been generally observed, and sometimes there is
strong debate \citep[e.g.,][]{mupesn2003,kodi2005}. We note that the
supercritical accretion explanation for ultralumineous very soft
sources might also need to involve IMBHs, instead of stellar-mass
BHs. This is because edge-on viewing is required to have very soft
spectra, by viewing the cool outflow only (not the central hot
region), but at high inclinations, it is hard to achieve
super-Eddington luminosities due to self-absorption and lack of
beaming effect \citep{ohmona2005}.

We include seven very soft sources here, all with $kT_{\rm BB}<0.2$
keV (see below) and at high Galactic latitudes ($b=-22\degr$ for
source \#65459 and $|b|\gtrsim 60\degr$ for the others). LWB12
selected out these sources for their low hardness ratios unseen in
most AGNs (Table~\ref{tbl:dercat}). To compare more with AGNs, we also
calculated the strength of soft excess $R_{\rm exc}$, defined as the
ratio of unabsorbed 0.3--2 keV fluxes in the cool and hard components
\citep{gido2004}, using the MCD+PL model (the BB+PL model gave
comparable results). Because our sources have no significant hard
X-ray emission, the PL component is generally not needed, in which
case we fixed $\Gamma_{\rm PL}$ at a value of 2 often seen in AGNs
(LWB12) and calculated the 90\% lower limit of $R_{\rm exc}$. Our
sources have $R_{\rm exc}\gtrsim6$ (see below), while most AGNs have
$R_{\rm exc}\lesssim1$ \citep{gido2004}.

\subsubsection{Two Sources Unassociated with Nearby Galaxies}
We first look at two sources, \#107102 and \#65459, which are not
associated with any galaxy in the Third Reference Catalog
\citep[RC3,][]{dedeco1991}. Source \#107102 is coincident with an
optical source from the Sloan Digital Sky Survey (SDSS, Release
8\footnote{http://www.sdss3.org/}, Table~\ref{tbl:sdssmat}), which
appears slightly extended. Its Magellan optical spectrum taken by
\citet{hokite2012} showed only narrow emission lines (no broad
H$_\alpha$ or H$_\beta$ lines) that are consistent with a low-mass
($10^5$ \msun) type 2 Seyfert nucleus at a redshift of 0.11871 (a
luminosity distance of 532 Mpc). The source was detected in all its
three \xmm\ observations, with $V_{\rm var14}$=2.1. Their spectra are
all very soft ($kT_{\rm BB}$$\sim$0.12--0.14 keV or $kT_{\rm
  MCD}$$\sim$0.15--0.18 keV, Table~\ref{tbl:spfit},
Figure~\ref{fig:srcspec}, $R_{\rm exc}>36$). The 0.3--10 keV absorbed
luminosity reached a peak of 3.6$\times$10$^{42}$ erg s$^{-1}$ in the
second observation (0306630101). Remarkably, the source showed large
coherent oscillations at a period about 3.8 hr in the two observations
in 2005 December but not in the observation in 2003 July
(Figure~\ref{fig:107102foldcurve}). However, the minima and maxima of
the light curves in the two observations in 2005 do not seem to be
well in phase (Figure~\ref{fig:107102foldcurve}). Therefore we have
probably detected a quasi-periodic oscialtion (QPO). The folded light
curves in two energy bands (0.2--0.8 keV and 0.8--2 keV) in
Figure~\ref{fig:107102foldcurve} indicate that the modulation seems
larger at higher energies. Considering the strong fast variability and
little neutral absorption in X-rays, the lack of H$_\alpha$ and
H$_\beta$ lines in the Magellan optical spectrum taken by
\citet{hokite2012} should indicate the real absence of the broad line
region, instead of being hidden (i.e., not a standard type 2 AGN),
similar to \object{GSN 069} \citep{misaro2013}. We have devoted a
separate work \citep{liirgo2013} to carry out detailed studies of the
source, including formal calculation of the QPO significance by
carefully modeling the red noise (the 99.9\% upper limits (dashed
lines) in Figure~\ref{fig:107102foldcurve} account for only the
Poisson noise but not the red noise) and the spectral fits with more
physical models. We refer to this work for more discussion on the
nature of the source. We note that the source was also discovered
independently by \citet{tekaaw2012}, who reported similar spectral and
timing properties of the source as we show above. One main difference
is that we show that the QPO is strong in both observations in 2005
December, not just in the first one as claimed by \citet{tekaaw2012}.

Source \#65459 was detected in one observation on 2005 September 30,
but not in an earlier observation on 2003 December 19, with $V_{\rm
  var14}$$>$3.2 (Figure~\ref{fig:srcltlc}). The detection has $kT_{\rm
  BB}$$=$0.11 keV or $kT_{\rm MCD}$$=$0.14 keV (Table~\ref{tbl:spfit},
Figure~\ref{fig:srcspec}, $R_{\rm exc}>12$). It shows some short-term
variability (Figure~\ref{fig:srclc}). The source has no counterpart
from the USNO-B1.0 Catalog or the 2MASS PSC
(Table~\ref{tbl:dercat}). It was not observed by the SDSS. Future deep
optical imaging is needed to check whether it is associated with any
galaxy to help to pin down its nature.

\subsubsection{Two Sources Near NGC 4151}
We next look at two sources, \#101369 and \#101335. They have angular
separations from the center of the galaxy NGC 4151 of
$\alpha$=7.1$\arcmin$ and 6.6$\arcmin$, respectively. They were not
considered to be associated with this galaxy by LWB12, because their
ratios of the angular separation to the $D_{25}$ isophote elliptical
radius $\alpha/R_{\rm 25}$ are 2.8 and 2.5, respectively, larger than
the threshold value of 2 used to determine the
association. Considering that NGC 4151 in fact has outer arms observed
to extend out to 6$\arcmin$ \citep{pehoax1992}, it is still possible
that both sources are in this galaxy. Alternatively, they could be
associated with distant galaxies, considering that both of them have
SDSS optical counterpart candidates (Table~\ref{tbl:sdssmat}).

Source \#101369 has eight \xmm\ observations spanning nearly six years
but was only detected in the first three observations which are
continuous in time around 2000 December 22 and have very similar
spectra, resulting in $V_{\rm var14}$$>$5
(Figure~\ref{fig:srcltlc}). The X-ray spectra are supersoft ($kT_{\rm
  BB}$$\sim$0.08 keV, Table~\ref{tbl:spfit}, Figure~\ref{fig:srcspec},
$R_{\rm exc}=11$). The maximum 0.3--10 keV absorbed luminosity is
1.3$\times$10$^{39}$ erg s$^{-1}$ if it is in NGC 4151 \citep[20.3
  Mpc,][]{libr2005}. The SDSS counterpart candidate of the source is
slightly extended (the $r$-band Petrosian radius is $\sim$1\farcs5)
and has a photometric redshift of 0.17$\pm$0.03, corresponding to a
luminosity distance of 790 Mpc. The maximum 0.3--10 keV absorbed
luminosity of the source is 2.0$\times$10$^{42}$ erg s$^{-1}$ using
this distance.

Source \#101335 was detected in all its eight \xmm\ observations
spanning nearly six years, with $V_{\rm var14}$$=$2.6
(Figure~\ref{fig:srcltlc}). We note that the source was at the CCD
edge in some observations, which could affect the calculation of the
long-term variability. The source was also detected in one {\it
  Chandra} detection in the CSC (Figure~\ref{fig:srcltlc}) and in
several occasions by {\it ROSAT}\ HRI and PSPC in the early 1990s with
similar fluxes \citep{whgian1994}, indicating the persistence of the
source over nearly two decades. The fits with an absorbed PL to band
count rates in \xmm\ observations indicate that the source spectra
were generally soft except the latest observation 0402660201 on 2006
November 29 (LWB12). We analyzed this observation and another
observation 0112830201 on 2000 December 22 in detail
(Tables~\ref{tbl:obslog} and \ref{tbl:spfit}, Figures~\ref{fig:srclc}
and \ref{fig:srcspec}). The second observation has a higher S/N, and
it cannot be fitted well with a BB, while the fit with a MCD is
acceptable, with $kT_{\rm MCD}=0.21$ keV (Table~\ref{tbl:spfit},
$R_{\rm exc}>6$). The observation 0112830201 is fainter and shorter
and has a low S/N, but the fit with a MCD requires a significantly
higher temperature $kT_{\rm MCD}=0.59$ keV, and a PL fit gives
$\Gamma_{\rm PL}$=2.1, which probably indicates a state transition. If
the source is in NGC 4151, the 0.3--10 keV absorbed luminosity reached
2.3$\times$10$^{39}$ erg s$^{-1}$ and 1.7$\times$10$^{39}$ erg
s$^{-1}$ in the soft observation 0112830201 and the hard observation
0402660201. The SDSS counterpart candidate of the source is faint and
appears blue and point-like (Table~\ref{tbl:sdssmat}).

Therefore, if they are in NGC 4151, both sources \#101369 and \#101335
would be very soft ULXs and could be due to cool thermal disk emission
from accreting IMBHs or outflows from supercritical accretion. For
source \#101335, there is possible state transition without large
variation in luminosity, which could possibly be explained more easily
with the former model. If the SDSS sources are their counterparts in
this galaxy, these counterparts would have absolute $r$-band
magnitudes about $-12.0$ and $-9.9$, respectively, which are very
bright and could be very massive star clusters. Alternatively, these
SDSS sources are still their counterparts but are distant
galaxies. Then sources \#101369 and \#101335 could be caused by
nuclear activity. Considering its possible transient nature, sources
\#101369 could be a TDE. For source \#101335, whose flux was
relatively constant over nearly two decades, it could be an ultrasoft
AGN similar to \object{GSN 069} \citep{misaro2013}.

\subsubsection{Three Sources Associated with Nearby Galaxies}
The other three very soft sources that we analyzed are sources \#8085,
\#183876, and \#187831, which LWB12 found to be likely within some RC3
galaxies. Source \#8085 was detected on the outskirts of NGC 300
($\alpha/R_{\rm 25}$=1.1 (Table~\ref{tbl:dercat})) in all its four
\xmm\ observations spaning 4.9 years, but exhibiting large long-term
variability ($V_{\rm var14}$=36, Figure~\ref{fig:srcltlc}). Clear
spectral change was observed, with $kT_{\rm BB}$ varying from 0.07 to
0.12 keV (Table~\ref{tbl:spfit}, $R_{\rm exc}>174$). Its maximum
0.3--10 keV absorbed luminosity is 1.2$\times$10$^{38}$ erg s$^{-1}$,
assuming a distance of 2 Mpc \citep{frmagi2001}.

Source \#183876 was detected toward Seyfert 1 galaxy NGC 7314
($\alpha/R_{\rm 25}$=0.8) in two \xmm\ observations separated by 6
years, with no significant spectral variation observed ($V_{\rm
  var14}$$=$1.2, Figure~\ref{fig:srcltlc}). The spectral fitting
results of one observation (on 2006 May 4) are given in
Table~\ref{tbl:spfit} and shown in Figure~\ref{fig:srcspec}. The
spectrum has $kT_{\rm BB}$$=$0.17 keV or $kT_{\rm MCD}$$=$0.21 keV
($R_{\rm exc}=11$). Although we added a weak PL in the fit, this
component can be due to contamination from the nucleus. There are also
two detections in the CSC about one month after the second
\xmm\ observation, with comparable fluxes
(Figure~\ref{fig:srcltlc}). Assuming a distance of 12.9 Mpc
\citep{libr2005}, its 0.3--10 keV absorbed luminosity is about
8.6$\times$10$^{38}$ erg s$^{-1}$.

Source \#187831 was detected in the second of its three
\xmm\ observations of Seyfert 2 galaxy NGC 7582 ($\alpha/R_{\rm
  25}$=0.9, $V_{\rm var14}$$>$5.0, Figure~\ref{fig:srcltlc}). The
detection has $kT_{\rm BB}$$=$0.12 keV or $kT_{\rm MCD}$$=$0.14 keV
(Table~\ref{tbl:spfit}, Figure~\ref{fig:srcspec}, $R_{\rm
  exc}>34$). The light curve exhibits significant variability
(Figure~\ref{fig:srclc}). Assuming a distance of 17.6 Mpc
\citep{libr2005}, we obtained a 0.3--10 keV absorbed luminosity of
1.1$\times$10$^{39}$ erg s$^{-1}$, making it a ULX. We note that this
source is also close to another galaxy \object{6dFGS g2318191-422235}
at a redshift of 0.05814 \citep{joresa2009}, at a separation of
4\farcs6 (corresponding to 5 kpc). Thus it is possible that source
\#187831 is in this galaxy instead, which would increase its 0.3--10
keV absorbed luminosity to be 2.3$\times$10$^{41}$ erg s$^{-1}$
(assuming a luminosity distance of 249.6 Mpc).

Sources \#183876 and \#187831 have maximum luminosities well above the
Eddington limit of WDs ($\sim$2$\times$$10^{38}$ erg s$^{-1}$) and are
probably due to cool thermal disk emission from IMBHs or outflows from
supercritical accretion. Source \#8085 could have a similar origin,
instead of nuclear burning of material accreted by a WD, for its
relatively high luminosity and BB temperature (0.12 keV) in one
observation. Among these three sources, only source \#8085 showed
clear spectral variation, which could be used to differentiate between
the thermal disk and outflow models. Both the BB and MCD fits suggest
significant temperature variation at a relatively constant emission
area (Table~\ref{tbl:spfit}), which seems to favor the thermal disk
model, but this should be confirmed with future high-quality
observations.

\subsection{Sources with Large Long-term Variability}
\subsubsection{Off-nuclear Extragalactic Source Candidates}
We analyzed six sources that appear in nearby galaxies and show large
long-term variability. Because of their high peak luminosities
($>10^{37}$ erg~s$^{-1}$, see below) and the lack of very hard spectra
expected for accretion-powered X-ray pulsars (LWB12), they are most
likely accreting BHs or weakly magnetized NSs (they are mostly
low-mass X-ray binaries) if they are indeed extragalactic (CVs have
luminosities $<10^{33}$ erg~s$^{-1}$ when they are not in the
supersoft phase \citep{muarba2004}). The differentiation between
accreting BHs and accreting weakly magnetized NSs is generally
difficult, but hints can still be obtained in some cases. Observation
of luminosities much higher than the Eddington limit for accreting NSs
(1.8$\times$10$^{38}$ erg s$^{-1}$ assuming a 1.4-\msun\ NS and the
pure hydrogen accreting material) and/or observation of soft spectra
of $kT_{\rm MCD}<1$ keV at high luminosities will favor the accreting
BH scenario \citep{remc2006}, because accreting weakly magnetized NSs,
even when they are in the soft state, seldom exhibit so soft spectra
due to presence of the hot emission from the impact of accreting
materials onto the NS surface
\citep{dogi2003,lireho2007,lireho2009,lireho2010,lireho2012}. NS X-ray
binaries can appear very soft only near quiescence \citep[$<$10$^{34}$
  erg s$^{-1}$,][]{weba2007}.

We first look at four sources (\#7204, \#29251, \#105130 and \#241580)
that are probably BH X-ray binaries. Source \#7204 was detected in one
of nine observations of NGC 253 ($\alpha/R_{\rm 25}$=0.74, $V_{\rm
  var14}$$>$23, Figure~\ref{fig:srcltlc}). The spectrum was fitted
with an absorbed PL with $\Gamma_{\rm PL}$=2.9 by
\citet{bagrko2008}. We reanalyzed the data and obtained a consistent
fit (Table~\ref{tbl:spfit}). Considering the high value of
$\Gamma_{\rm PL}$, we also fitted the spectra with a MCD plus a weak
PL ($\Gamma_{\rm PL}$ was fixed at a value of 2 due to its large
uncertainty). The best-fitting disk temperature is about 0.5 keV, and
the 0.3--10 keV absorbed luminosity is 0.86$\times$10$^{38}$ erg
s$^{-1}$ if a distance of 3.0 Mpc is assumed \citep{li2011}.

Source \#29251 was detected in two of fourteen observations of NGC
1313 ($\alpha/R_{\rm 25}$=0.18, $V_{\rm var14}$$>$30,
Figure~\ref{fig:srcltlc}). These two detections, separated by only 76
days, are probably from the same outburst. The fluxes and spectral
shapes are similar in these two detections. We fitted the spectrum of
one observation using the MCD model (Table~\ref{tbl:spfit},
Figure~\ref{fig:srcspec}) and obtained $kT_{\rm MCD}$=0.8 keV and a
0.3--10 keV absorbed luminosity of 4.0$\times$10$^{38}$
erg~s$^{-1}$, assuming a distance of 3.7 Mpc \citep{libr2005}. We note
that an absorbed PL with $\Gamma_{\rm PL}=2.4$ can also fit the
spectrum (Table~\ref{tbl:spfit}). The source was also detected by {\it
  ROSAT}, with a slightly higher luminosity than we report here
\citep{libr2005}, indicating the recurrent nature of the outbursts.

Source \#105130 was detected in three of five observations on the
outskirts of NGC 4395 ($\alpha/R_{\rm 25}$=1.65, $V_{\rm
  var14}$$>$65). We analyzed the brightest observation, taken on 2004
July 3, and found that the MCD model ($kT_{\rm MCD}$=0.6 keV) fitted
the spectrum better than the PL and BB models, with the $\chi^2$
values decreased by 39 and 181 for 138 degrees of freedom,
respectively (Table~\ref{tbl:spfit}, Figure~\ref{fig:srcspec}). No
significant short-term variation is seen in this observation
(Figure~\ref{fig:srclc}). Assuming the source to be in NGC 4395 at a
distance of 3.6 Mpc \citep{libr2005}, we obtained the maximum
0.3--10 keV absorbed luminosity of 1.9$\times$10$^{38}$ erg
s$^{-1}$. The source has two detections in the CSC
(Figure~\ref{fig:srcltlc}), which, in addition to \xmm\ observations,
seem to suggest at least three outbursts within four years. The source
was also detected in one {\it ROSAT}\ HRI observation 1RH702725N00 in
1996 \citep[2.3$\times10^{38}$ erg s$^{-1}$,][]{libr2005}.

Source \#241580 was detected in two observations in 2008 May--June in
the direction of NGC 4490 ($\alpha/R_{\rm 25}$=0.87), but not in
another observation in 2002 May ($V_{\rm var14}$$>$16). The spectrum
of the brightest observation can be fitted with an absorbed PL
(Table~\ref{tbl:spfit}, Figure~\ref{fig:srcspec}), with the
corresponding 0.3--10 keV absorbed luminosity of 7.6$\times$10$^{38}$
erg s$^{-1}$ if a distance of 7.8 Mpc is assumed \citep{li2011}.

We see that sources \#7204, \#29251, and \#105130 have soft spectra
with $kT_{\rm MCD}\lesssim 0.8$ keV and that sources \#29251, \#105130
and \#241580 have maximum 0.3--10 keV absorbed luminosities above the
Eddington limit for accreting NSs. Therefore, these four sources are
probably BH X-ray binaries.

We next look at two sources (\#222148 and \#222243) in M 31. Source
\#222148 was detected in five of eighteen observations of M 31
($\alpha/R_{\rm 25}$=0.15, $V_{\rm var14}$$>$152). There are also
eleven detections in the CSC. All these \xmm\ and {\it Chandra}
detections spanned about 407 days from 2007 November to 2008 December
(Figure~\ref{fig:srcltlc}). The outburst was discovered by
\citet{gagast2007} from a 5ks {\it Chandra} observation on 2007 July
31, indicating that the outburst actually began earlier. To check
whether the outburst is still ongoing, we analyzed 19 new observations
that were not included in the 2XMMi-DR3 catalog but publicly available
as of 2012 August (so covering data before 2011 July). The spectral
fits using an absorbed PL for these 19 observations, in addition to
the five detections included in the 2XMMi-DR3 catalog, are given in
Table~\ref{tbl:spfit}. The results show that the source has been X-ray
bright for more than four years (this is also supported by
\citet{hopihe2013}, who recently analyzed {\it Chandra} HRC
observations of the source). The fluxes varied strongly at the
beginning of the outburst and became fairly constant later. The
spectra are consistently hard, with $\Gamma_{\rm PL}$$\sim$1.8. One
sample spectral fit is shown in Figure~\ref{fig:srcspec}, using
observation 0672130101, which has the most counts. The maximum 0.3--10
keV absorbed luminosity is 6$\times$10$^{37}$ erg s$^{-1}$, and the
average is around 4$\times$10$^{37}$ erg s$^{-1}$ \citep[assuming a
  distance of 780 kpc,][]{stga1998}. Although we found no optical and
IR counterparts from the USNO-B1.0 Catalog and 2MASS PSC, there is an
optical source \object{LGGS J004211.09+410431.0} in \citet{maolho2007}
at a separation of 1\farcs8, which has a $V$-band magnitude of
22.4$\pm$0.1 and colors of $B-V=1.4\pm0.4$, $V-R=0.6\pm0.1$ and
$R-I=1.0\pm0.1$ and is probably a late-type dwarf star.

There have been some X-ray binaries known to remain X-ray bright most
of the time since they were discovered to begin the outburst, e.g.,
the BH X-ray binary candidates \object{Swift J1753.5-0127}
\citep[since 2005 May, ][]{sofetu2010} and \object{GRS 1915+105}
\citep[since 1992 August,][]{mcre2006} and the accretion-powered
millisecond X-ray pulsar \object{HETE J1900.1-2455} \citep[since 2005
  June,][]{gamokr2007}. \object{Swift J1753.5-0127} is essentially in
the hard state, with 2--10 keV luminosity mostly around
5$\times$10$^{36}$ erg s$^{-1}$ \citep{sofetu2010}. \object{HETE
  J1900.1-2455} has a similar mean luminosity
\citep[4.4$\times$$10^{36}$ erg s$^{-1}$ in 2.5--25
  keV,][]{gamoch2008} and is mostly (but not always) in the hard state
based on the hardness ratios that we calculated from the {\it RXTE}
monitoring pointed observations (following a method as described in
\citet{lireho2007}). \object{GRS 1915+105} exhibits complicate
behavior and has shown several states (including both the hard and
soft/thermal states). If source \#222148 is a BH X-ray binary, it
should be in the hard state. To check whether the source could be an
accreting weakly magnetized NS in the hard state, in which a BB
component is normally present and contributes a few percent (but
$<$10\%) of the luminosity \citep{lireho2007,lireho2010}, we added a
BB component in the fit to the observation 0672130101. We found that
the spectral fit was not significantly improved (the $\chi^2$
decreased by 2.6 from 63.0 while the degrees of freedom decreased by 2
from 74), with the BB flux estimated to be $<$20\% of the total
(0.3--10 keV, absorbed) at a 90\%-confidence level. Thus, although we
do not need the BB in the fits, we cannot exclude its presence at the
level expected for accreting weakly magnetized NSs in the hard
state. However, the luminosity of accreting weakly magnetized NSs in
the hard state is normally lower than 10\% of the Eddington limit
\citep{lireho2007,lireho2010,gamoch2008,hawach2009}, while source
\#222148 is slightly brighter (by a factor of 2 or more if the
luminosity is estimated over a broader energy band). We note that we
cannot exclude the possibility that the source is in the soft state of
an accreting weakly magnetized NS based on the spectral fit. To
demonstrate this, we also fitted the spectrum of observation
0672130101 with a model of MCD+BB, which are often used to fit the
soft state of an accreting weakly magnetized NS
\citep{miinko1984,lireho2007,lireho2009,lireho2010,lireho2012} and
obtained a similar quality of the fit as that using an absorbed PL
(Table~\ref{tbl:spfit}). This is because the soft-state spectra of
accreting weakly magnetized NSs can still appear hard below about 7
keV. Future broad band observations with observatories such as {\it
  NuSTAR} can break such a model degeneracy easily.

Source \#222243 was detected in three of nineteen observations of M
31. It is only $\alpha$=1.5$\arcmin$ ($\alpha/R_{\rm 25}$=0.02) away
from the center of M 31. LWB12 measured $V_{\rm var14}$$>$6 and
$V_{\rm var}$(0.2-12.0 keV)$>$10, but these variation factors were
most likely underestimated due to the bright diffuse emission in the
galaxy center. The CSC indicates that the source was also detected in
nine {\it Chandra} observations from 2007 March to 2008 December, with
0.1--10 keV fluxes about 10$^{-14}$ erg~s$^{-1}$~cm$^{-2}$. In fact,
there are many more {\it Chandra} observations of this source,
indicating that the source is generally faint for 12 years
\citep{bagamu2012,hopihe2013}. In comparison, the three
\xmm\ observations in which the source was detected were made at the
beginning of 2007 January, with 0.2--12.0 keV fluxes in the range of
(1.4--12.2)$\times$10$^{-13}$ erg~s$^{-1}$~cm$^{-2}$. Their spectra
are all hard. We analyzed the brightest observation 0405320801 and
found clear residuals when we fitted the spectrum with a single
component model (i.e., an absorbed MCD, BB, or PL). We obtained an
acceptable fit with a MCD+BB model (Table~\ref{tbl:spfit} and
Figure~\ref{fig:srcspec}), indicating that the source might be an
accreting weakly magnetized NS in the soft state (see discussion above
for source \#222148). The 0.3--10 keV absorbed luminosity is
6.8$\times$10$^{37}$ erg s$^{-1}$ \citep[assuming a distance of 780
  kpc,][]{stga1998}. The source is probably in the globular cluster
candidate BH16 \citep[i.e., PB-in7, ][]{gafebe2004,bagamu2012}.

\subsubsection{Galactic Source Candidates}

Sources \#142052, \#165335 and \#174295 are at low Galactic latitudes
($|b|<3\degr$) and have Galactic longitudes of about $327\degr$,
$11\degr$, and $80\degr$, respectively. They are highly variable
($V_{\rm var14}>20$, Figure~\ref{fig:srcltlc},
Table~\ref{tbl:dercat}). Their X-ray spectra are generally hard and
can be fitted with an absorbed PL (when detected,
Table~\ref{tbl:spfit}, Figure~\ref{fig:srcspec}). They all have
optical and IR counterparts found within 1.3$\arcsec$ from the
USNO-B1.0 Catalog and the 2MASS PSC. LWB12 classified them as compact
object systems based on high X-ray-to-IR flux ratios, lack of flares,
and large variability.

Source \#142052 was detected in all four \xmm\ observations and one
{\it Chandra} observation (ObsID: 7287, from the CSC) spanning five
years, with $V_{\rm var14}$=23 (Figure~\ref{fig:srcltlc}). It is
probably \object{AX J1549.8-5416}, which was detected in the {\it
  ASCA} Galactic Plane Survey \citep{sumika2001}. Thus the source is
probably persistent but highly variable. In the spectral fits to two
long \xmm\ observations (0402910101 and 0560181101), a Gaussian Fe
line was detected above 3$\sigma$ (based on the normalization), with
an equivalent width (EW) of about 1.0 keV (Table~\ref{tbl:spfit}). The
source peak luminosity is in the range between about
2$\times$10$^{32}$ and 10$^{34}$ erg s$^{-1}$ (0.3--10 keV, absorbed)
if the source distance is between 1--8.5 kpc. Its optical counterpart
is probably NSV 20407, a variable star with the $B$-band magnitude
ranging from $<18$ to 16.7 \citep{sadu2009}, supporting to identify it
as a CV.  Further considering that the magnetic CVs often exhibit the
Fe emission line \citep{ezis1999}, source \#142052 is probably one
such system.

Source \#165335 was detected in one of its two \xmm\ observations
($V_{\rm var14}>25$). It was also detected in a {\it Chandra}
observation in 2006 (ObsID: 6405, from the CSC) and is source 23 in
\citet{prtssu2009} in the direction of the MRR 32 X-ray cluster. The
flux in this {\it Chandra} observation is very low, resulting in a
0.5--7 keV flux variation factor of 167, compared with the
\xmm\ detection. The 0.3--10 keV absorbed luminosity in the
\xmm\ detection is within the range of 8$\times$$10^{31}$ to
6$\times$$10^{33}$ erg s$^{-1}$ if the source distance is between
1--8.5 kpc. Considering that it appears in a star forming region, the
bright \xmm\ observation, which last 32 ks and showed no obvious
variability, might be due to an extremely long (probably several days)
stellar X-ray flare. However, stellar X-ray flares with durations of
days and flux increasing factors of more than one hundred are rarely
seen. Flares lasting for days or more have been detected before from
RS CVn stars \citep[e.g.,][and references
  therein]{kusc1996,enstku1997}, but they have flux increasing factors
of only a few tens. LWB12 reported several flares (e.g., sources
\#64367 and \#146561) with flux increasing factors of more than one
hundred, but they remained bright for only a few ks. If it is not a
star, it is most probably a CV or a very faint X-ray transient (VFXT)
with a BH or NS as the accretor \citep{wiinru2006}, considering the
very low peak luminosity of the source. VFXTs are transient sources
with peak outburst luminosities between $10^{34}$ and
$\lesssim$$10^{36}$ erg s$^{-1}$. The lower limit of $10^{34}$ erg
s$^{-1}$ is adopted in the definition so that they are not CVs but
probably accreting BHs or NSs. It is not clear whether they can in
fact have fainter outbursts. Most known VFXTs were detected toward the
Galactic center from the intensive monitoring of the Galactic center
by {\it XMM-Newton} and {\it Chandra}
\citep{sawade2005,mupfba2005,wiinru2006}. Our knowledge on this class
of objects is still very limited.

Source \#174295 was securely detected in an observation on 2004
November 9, with a hard and highly absorbed spectrum as indicated by
the fits with an absorbed PL (Table~\ref{tbl:spfit},
Figure~\ref{fig:srcspec}), but it was hardly detected in the
observations ten days before and ten days after it, indicating a short
outburst. The source has a candidate SDSS point-like counterpart
(Table~\ref{tbl:sdssmat}), which appears very red, probably caused by
strong absorption. The maximum 0.3--10 keV absorbed luminosity would
be between 2$\times$$10^{32}$ and $10^{34}$ erg~s$^{-1}$ if the source
distance is within 1--8.5 kpc. Similar to our discussion for source
\#165335 above, this source might be a CV, a VFXT with a BH or NS as
the accretor, or a star with an extremely energetic stellar flare.

\section{CONCLUSIONS}
\label{sec:conclusion}
We have studied 18 sources from the 2XMMi-DR3 catalog that were poorly
studied in the literature but show interesting properties of
periodicity, very soft spectra and/or large long-term variability in
X-rays in order to investigate their nature. Our findings can be
summarized as follows:
\begin{enumerate}
\item
Two sources have been persistently detected with hard spectra
($\Gamma_{\rm PL}\sim 1$). One of them (source \#150140) was detected
on the outskirts of the young open cluster NGC 6231 and showed a 1.62
hr periodicity in six \xmm\ observations and one {\it Chandra}
observation separated by nearly four years. There is also a $\sim$2.1
hr period candidate in the other source (\#122979), but it needs to be
confirmed with future long observations. Both sources are good
magnetic CV candidates.
\item
Seven very soft sources ($kT_{\rm BB}<0.2$ keV) are probably in other
galaxies, which would imply the luminosities ranging from about
$10^{38}$ to $10^{42}$ erg s$^{-1}$. Some of them might be coincident
with galactic nuclei and best explained as ultrasoft AGNs or TDEs,
while the others are probably off-nuclear and could be cool thermal
disk emission from accreting IMBHs or edge-on viewing of optically
thick cool outflows from supercritical accretion. One source
(\#107102) is an AGN candidate and showed an intermittent QPO at about
3.8 hr, i.e., in two observations in 2005 December but not in one
observation in 2003 July.
\item
Six highly variable sources with spectra harder than the above very
soft sources appear in nearby RC3 galaxies and probably have
luminosities above $10^{37}$ erg s$^{-1}$, making them great
candidates for extragalactic X-ray binaries. One source (\#222148, in
M 31) has remained X-ray bright and hard in 0.3--10 keV since it was
discovered to enter an outburst on 2007 July 31 and thus is probably a
new-born persistent source.
\item
Three highly variable hard sources appear at low galactic latitudes
and have maximum luminosities below about $10^{34}$ erg s$^{-1}$ if
they are in our Galaxy. They could be CVs, VFXTs with a BH or NS as
the accretor, or stars with extremely strong flares.
\end{enumerate}

Although most of our sources still require future optical
spectroscopic observations and/or long-term X-ray monitoring to
finally pin down their nature, it is clear that our sample contains a
variety of objects with all kinds of spectral behavior. Some of them
belong to rare classes that still have relatively small numbers of
sources known, such as very soft X-ray sources, very faint transient
and new-born persistent sources, and our discovery of new objects for
these classes is thus important for our understanding of them. We
found their properties mostly consistent with previous studies, but
some new results were also obtained. Focusing on the seven very soft
X-ray sources, we found the existence of both supersoft and quasisoft
X-ray sources and the possibility that some of them are the same class
of objects because supersoft and quasisoft spectra can be observed in
a single source (see source \#8085), consistent with previous studies
\citep[e.g.,][]{kodi2005}. However, we also observed some evidence
that they displayed various kinds of spectral evolution behavior
(being transient or persistent with either small or large long-term
variability and displaying state transition) often seen in X-ray
binaries. It is not clear whether this is completely because our
sample might include several classes of objects. Our discovery of some
unique objects is also significant, especially source \#107102, as an
AGN candidate with very soft spectra and a strong QPO in X-rays
\citep{liirgo2013}. There is only one other AGN known to show
similarly weak hard X-ray emission (\object{GSN 069},
\citealt{misaro2013}) and only two other galactic nuclei with QPOs
significantly detected (\object{RE J1034+396}, \citealt{gimiwa2008};
\object{Swift J164449.3+573451}, \citealt{remire2012}).

Acknowledgments: We thank the anonymous referee for the helpful
comments. We acknowledge the use of public data from the {\it Chandra}
and \xmm\ data archives, the 2XMM Serendipitous Source Catalog,
constructed by the XMM-Newton Survey Science Center on behalf of ESA,
and the {\it Chandra} Source Catalog, provided by the {\it Chandra}
X-ray Center as part of the {\it Chandra} Data Archive.

\clearpage
\LongTables
\tabletypesize{\tiny}
\setlength{\tabcolsep}{0.07in}
\begin{deluxetable*}{r|c|c|cc|cc|cc|c|cc}
\tablecaption{Spectral fit results \label{tbl:spfit}}
\tablewidth{0in}
\tablehead{SRCID& Obs ID & $N_{\rm H}$ &  $kT_{\rm MCD}$ &$N_{\rm MCD}$ &  
  $kT_{\rm BB}$ &$N_{\rm BB}$ & 
  $\Gamma_{\rm PL}$ &$N_{\rm PL}$ & 
   $\chi^2_\nu(\nu)$\tablenotemark{a} & $F_{\rm abs}$ & $F_{\rm unabs}$\\
               &  & (10$^{22}$ cm$^{-2}$)&
  (keV) & &
  (keV) & &
        & ($10^{-6}$)&
   & \multicolumn{2}{c}{(10$^{-13}$ erg s$^{-1}$ cm$^{-2}$)}\\
(1)&(2)&(3)&(4)&(5)&(6)&(7)&(8)&(9)&(10)&(11)&(12)
}
\startdata
\multirow{2}{*}{7204} & \multirow{2}{*}{0152020101} & $0.01^{+0.02}_{-0.01}$  & $ 0.47^{+ 0.04}_{-0.05}$  & $0.06^{+0.04}_{-0.02}$ &\nodata&\nodata& $ 2.0$ & $ 4^{+ 3}_{-1}$ &$0.91(101)$ & $ 0.80^{+ 0.06}_{-0.06}$  & $ 0.82^{+ 0.10}_{-0.08}$ \\
\cline{3-12}
 & & $0.17^{+0.03}_{-0.03}$ &\nodata&\nodata&\nodata&\nodata& $ 3.1^{+ 0.2}_{-0.2}$  & $39^{+ 5}_{-4}$ &$1.04(102)$ & $ 0.79^{+ 0.05}_{-0.05}$  & $ 2.1^{+ 0.4}_{-0.3}$ \\
\hline
\multirow{4}{*}{8085} & \multirow{2}{*}{0305860401}\tablenotemark{b} & $0.13^{+0.10}_{-0.07}$ &\nodata&\nodata& $0.12^{+0.02}_{-0.02}$  & $431^{+1275}_{-286}$ &\nodata&\nodata&$1.22( 20)$ & $ 2.6^{+ 0.2}_{-0.2}$  & $ 7^{+ 8}_{-3}$ \\
\cline{3-12}
& & $0.20^{+0.11}_{-0.08}$  & $ 0.14^{+ 0.02}_{-0.02}$  & $432^{+1874}_{-322}$ &\nodata&\nodata&\nodata&\nodata&$1.40( 20)$ & $ 2.5^{+ 0.2}_{-0.2}$  & $13^{+13}_{-6}$ \\
\cline{2-12}
     & \multirow{2}{*}{0112800201} & $0.13$&\nodata&\nodata& $0.07^{+0.01}_{-0.01}$  & $1092^{+1336}_{-574}$ &\nodata&\nodata&\nodata& $ 0.14^{+ 0.03}_{-0.02}$  & $ 0.8^{+ 0.2}_{-0.2}$ \\
\cline{3-12}
& & $0.20$ & $ 0.07^{+ 0.01}_{-0.01}$  & $2782^{+4467}_{-1650}$ &\nodata&\nodata&\nodata&\nodata&\nodata & $ 0.13^{+ 0.03}_{-0.02}$  & $ 1.8^{+ 0.5}_{-0.4}$ \\
\hline
\multirow{2}{*}{29251} & \multirow{2}{*}{0205230501} & $0.09^{+0.05}_{-0.04}$  & $ 0.8^{+ 0.1}_{-0.1}$  & $0.04^{+0.04}_{-0.02}$ &\nodata&\nodata&\nodata&\nodata&$0.68( 29)$ & $ 2.4^{+ 0.3}_{-0.3}$  & $ 3.0^{+ 0.4}_{-0.4}$ \\
\cline{3-12}
&& $0.28^{+0.08}_{-0.08}$ &\nodata&\nodata&\nodata&\nodata& $ 2.4^{+ 0.3}_{-0.3}$  & $131^{+41}_{-31}$ &$0.66( 29)$ & $ 3.1^{+ 0.5}_{-0.5}$  & $ 6^{+ 2}_{-1}$ \\
\hline
\multirow{2}{*}{65459} & \multirow{2}{*}{0306560301} & $0.00^{+0.06}$ &\nodata&\nodata& $0.11^{+0.01}_{-0.01}$  & $  10^{+  4}_{ -3}$ &\nodata&\nodata&$1.19( 22)$ & $ 0.11^{+ 0.01}_{-0.01}$  & $ 0.11^{+ 0.01}_{-0.01}$ \\
\cline{3-12}
& & $0.01^{+0.07}_{-0.01}$  & $ 0.14^{+ 0.02}_{-0.03}$  & $3^{+17}_{-2}$ &\nodata&\nodata&\nodata&\nodata&$1.06( 22)$ & $ 0.11^{+ 0.02}_{-0.02}$  & $ 0.14^{+ 0.13}_{-0.03}$ \\
\hline
\multirow{4}{*}{101335} & \multirow{2}{*}{0112830201} & $0.00^{+0.01}$ &\nodata&\nodata& $0.15^{+0.01}_{-0.01}$  & $10^{+2}_{-2}$ &\nodata&\nodata&$1.34( 52)$ & $ 0.44^{+ 0.03}_{-0.03}$  & $ 0.44^{+ 0.03}_{-0.03}$ \\
\cline{3-12}
                              &       & $0.00^{+0.01}$  & $ 0.21^{+ 0.01}_{-0.01}$  & $1.8^{+0.5}_{-0.4}$ &\nodata&\nodata&\nodata&\nodata&$0.92( 52)$ & $ 0.47^{+ 0.03}_{-0.03}$  & $ 0.47^{+ 0.03}_{-0.03}$  \\
\cline{2-12}
       &  \multirow{2}{*}{0402660201} & $0.0$&\nodata&\nodata&\nodata& \nodata& $ 2.1^{+ 0.3}_{-0.3}$  & $ 6^{+ 2}_{-2}$ &\nodata & $ 0.3^{+ 0.1}_{-0.1}$  & $ 0.3^{+ 0.1}_{-0.1}$ \\
\cline{3-12}
 & & $0.0$ & $ 0.6^{+ 0.2}_{-0.1}$  & $0.010^{+0.016}_{-0.007}$ &\nodata&\nodata&\nodata&\nodata& \nodata& $ 0.24^{+ 0.07}_{-0.06}$  & $ 0.24^{+ 0.07}_{-0.06}$ \\
\hline
\multirow{2}{*}{101369} & \multirow{2}{*}{0112830201} & $0.00^{+0.02}$ &\nodata&\nodata & $0.08^{+0.01}_{-0.01}$  & $146^{+119}_{-62}$  & $  2.0$ & $  0.8^{+ 0.4}_{-0.4}$ &$0.86( 22)$ & $ 0.28^{+ 0.03}_{-0.03}$  & $ 0.28^{+ 0.03}_{-0.03}$ \\
\cline{3-12}
                  & & $0.00^{+0.02}$  & $ 0.10^{+ 0.01}_{-0.01}$  & $56^{+58}_{-28}$ &\nodata&\nodata & $  2.0$ & $  0.8^{+  0.4}_{ -0.4}$ &$0.74( 22)$ & $ 0.28^{+ 0.03}_{-0.03}$  & $ 0.28^{+ 0.03}_{-0.03}$ \\
\hline
105130 & 0200340101 & $0.00^{+0.01}$  & $ 0.60^{+ 0.03}_{-0.03}$  & $0.05^{+0.01}_{-0.01}$ &\nodata&\nodata&\nodata&\nodata&$0.94(138)$ & $ 1.21^{+ 0.08}_{-0.07}$  & $ 1.21^{+ 0.08}_{-0.07}$ \\
\hline
\multirow{6}{*}{107102} & \multirow{2}{*}{0145800101} & $0.00^{+0.01}_{-0.00}$ &\nodata&\nodata & $0.12^{+0.01}_{-0.01}$  & $32^{+7}_{-6}$ &\nodata&\nodata&$1.19( 73)$ & $ 0.48^{+ 0.03}_{-0.03}$  & $ 0.48^{+ 0.03}_{-0.03}$ \\
\cline{3-12}
& & $0.00^{+0.03}_{-0.00}$  & $ 0.16^{+ 0.01}_{-0.02}$  & $8^{+9}_{-2}$ &\nodata&\nodata&\nodata&\nodata&$1.10( 73)$ & $ 0.50^{+ 0.03}_{-0.04}$  & $ 0.51^{+ 0.15}_{-0.03}$ \\
\cline{2-12}
&\multirow{2}{*}{0306630101} & $0.00^{+0.01}_{-0.00}$ &\nodata&\nodata & $0.14^{+0.01}_{-0.01}$  & $30^{+4}_{-3}$ &\nodata&\nodata&$1.09(142)$ & $ 1.05^{+ 0.04}_{-0.04}$  & $ 1.05^{+ 0.04}_{-0.04}$ \\
\cline{3-12}
& & $0.03^{+0.02}_{-0.02}$  & $ 0.18^{+ 0.01}_{-0.01}$  & $11^{+6}_{-3}$ &\nodata&\nodata&\nodata&\nodata&$1.04(142)$ & $ 1.06^{+ 0.04}_{-0.04}$  & $ 1.37^{+ 0.21}_{-0.17}$ \\
\cline{2-12}
&\multirow{2}{*}{0306630201} & $0.00^{+0.02}_{-0.00}$ &\nodata&\nodata & $0.13^{+0.01}_{-0.01}$  & $37^{+4}_{-4}$ &\nodata&\nodata&$1.05(144)$ & $ 0.77^{+ 0.03}_{-0.03}$  & $ 0.77^{+ 0.03}_{-0.03}$ \\
\cline{3-12}
& & $0.03^{+0.02}_{-0.02}$  & $ 0.15^{+ 0.01}_{-0.01}$  & $19^{+11}_{-7}$ &\nodata&\nodata&\nodata&\nodata&$1.04(144)$ & $ 0.76^{+ 0.03}_{-0.03}$  & $ 1.10^{+ 0.20}_{-0.15}$ \\
\hline
\multirow{2}{*}{122979}&0105261401 & \multirow{2}{*}{$0.00^{+0.05}$} &\nodata&\nodata&\nodata&\nodata & $ 0.9^{+ 0.2}_{-0.2}$  & $15^{+ 2}_{-2}$ &$1.07( 15)$ & $ 2.7^{+ 0.5}_{-0.5}$  & $ 2.7^{+ 0.5}_{-0.5}$ \\
\cline{2-2} \cline{4-12}
&0105261701 & &\nodata&\nodata&\nodata&\nodata & $ 1.0^{+ 0.3}_{-0.3}$  & $10^{+ 3}_{-3}$ &$0.60(  6)$ & $ 1.6^{+ 0.6}_{-0.5}$  & $ 1.6^{+ 0.6}_{-0.5}$ \\
\hline
\multirow{3}{*}{142052} &0203910101 & $0.42^{+0.07}_{-0.06}$ &\nodata&\nodata&\nodata&\nodata & $  1.6^{+  0.1}_{ -0.1}$  & $257^{+ 46}_{-39}$ &$0.95( 65)$ & $13^{+ 1}_{-1}$  & $19^{+ 1}_{-1}$ \\
\cline{2-12}
&0402910101\tablenotemark{c} & $0.41^{+0.04}_{-0.04}$ &\nodata&\nodata&\nodata&\nodata & $  1.7^{+  0.1}_{ -0.1}$  & $123^{+ 12}_{-11}$ &$1.11(215)$ & $ 6.3^{+ 0.3}_{-0.3}$  & $ 8.8^{+ 0.4}_{-0.4}$ \\
\cline{2-12}
&0410581901 & $0.41$&\nodata&\nodata&\nodata&\nodata & $  2.3^{+  0.3}_{ -0.3}$  & $ 23^{+  5}_{ -5}$ &\nodata & $ 0.5^{+ 0.2}_{-0.1}$  & $ 1.1^{+ 0.2}_{-0.2}$ \\
\cline{2-12}
&0560181101\tablenotemark{d} & $0.48^{+0.05}_{-0.04}$ &\nodata&\nodata&\nodata&\nodata & $  1.9^{+  0.1}_{ -0.1}$  & $130^{+ 15}_{-14}$ &$1.00(188)$ & $ 5.0^{+ 0.3}_{-0.3}$  & $ 8.0^{+ 0.5}_{-0.5}$ \\
\hline
&0109490101 & &\nodata&\nodata&\nodata&\nodata & $ 1.0^{+ 0.3}_{-0.3}$  & $14^{+ 6}_{-4}$ &$0.85( 14)$ & $ 2.0^{+ 0.4}_{-0.4}$  & $ 2.3^{+ 0.4}_{-0.4}$ \\
\cline{2-2} \cline{4-12}
\multirow{7}{*}{150140} &0109490201 & \multirow{6}{*}{$0.6^{+0.2}_{-0.2}$} &\nodata&\nodata&\nodata&\nodata & $ 1.3^{+ 0.3}_{-0.3}$  & $29^{+14}_{-10}$ &$0.87(  8)$ & $ 2.6^{+ 0.7}_{-0.6}$  & $ 3.2^{+ 0.7}_{-0.6}$ \\
\cline{2-2} \cline{4-12}
&0109490301 & &\nodata&\nodata&\nodata&\nodata & $ 0.7^{+ 0.2}_{-0.2}$  & $13^{+ 5}_{-4}$ &$0.84( 20)$ & $ 2.9^{+ 0.7}_{-0.6}$  & $ 3.2^{+ 0.6}_{-0.6}$ \\
\cline{2-2} \cline{4-12}
&0109490401 & &\nodata&\nodata&\nodata&\nodata & $ 0.9^{+ 0.3}_{-0.2}$  & $15^{+ 6}_{-4}$ &$0.69( 16)$ & $ 2.3^{+ 0.5}_{-0.5}$  & $ 2.6^{+ 0.5}_{-0.5}$ \\
\cline{2-2} \cline{4-12}
&0109490501 & &\nodata&\nodata&\nodata&\nodata & $ 1.0^{+ 0.3}_{-0.3}$  & $13^{+ 6}_{-4}$ &$1.04( 13)$ & $ 1.8^{+ 0.5}_{-0.4}$  & $ 2.1^{+ 0.5}_{-0.5}$ \\
\cline{2-2} \cline{4-12}
&0109490601 & &\nodata&\nodata&\nodata&\nodata & $ 0.8^{+ 0.2}_{-0.2}$  & $12^{+ 5}_{-3}$ &$0.65( 17)$ & $ 2.1^{+ 0.5}_{-0.4}$  & $ 2.4^{+ 0.5}_{-0.4}$ \\
\cline{2-12}
&6291 (CXO) & $0.8^{+0.3}_{-0.3}$ &\nodata&\nodata&\nodata&\nodata & $ 1.2^{+ 0.3}_{-0.3}$  & $30^{+18}_{-11}$ &$0.53( 18)$ & $ 3.1^{+ 0.5}_{-0.5}$  & $ 3.8^{+ 0.4}_{-0.4}$ \\
\hline
165335 & 0149610401 & $0.60^{+0.07}_{-0.06}$ &\nodata&\nodata&\nodata&\nodata& $ 2.1^{+ 0.1}_{-0.1}$  & $251^{+39}_{-33}$ &$1.04(108)$ & $ 6.8^{+ 0.5}_{-0.5}$  & $13^{+ 2}_{-1}$ \\
\hline
174295 &0200450301 & $1.3^{+0.3}_{-0.2}$ &\nodata&\nodata&\nodata&\nodata & $  1.9^{+  0.2}_{ -0.2}$  & $482^{+194}_{-131}$ &$1.16( 50)$ & $15^{+ 2}_{-2}$  & $29^{+ 7}_{-4}$ \\
\hline
\multirow{2}{*}{183876} & \multirow{2}{*}{0311190101} & $0.00^{+0.06}$ &\nodata&\nodata& $0.17^{+0.01}_{-0.02}$  & $  3^{+  5}_{ -1}$  & $  1.3^{+  1.0}_{ -0.4}$  & $ 2^{+ 2}_{-1}$ &$1.19( 46)$ & $ 0.43^{+ 0.08}_{-0.07}$  & $ 0.44^{+ 0.13}_{-0.07}$ \\
\cline{3-12}
& & $0.06^{+0.06}_{-0.04}$  & $ 0.21^{+ 0.04}_{-0.03}$  & $2^{+4}_{-1}$ &\nodata&\nodata & $1.3^{+  0.9}_{ -0.8}$  & $  2^{+  2}_{ -1}$ &$1.21( 46)$ & $ 0.43^{+ 0.08}_{-0.07}$  & $ 0.61^{+ 0.24}_{-0.14}$ \\
\hline
\multirow{2}{*}{187831} & \multirow{2}{*}{0204610101} & $0.00^{+0.03}$ &\nodata&\nodata& $0.12^{+0.01}_{-0.01}$  & $ 21^{+ 20}_{ -4}$ &\nodata&\nodata&$1.24( 68)$ & $ 0.31^{+ 0.02}_{-0.02}$  & $ 0.31^{+ 0.10}_{-0.02}$ \\
\cline{3-12}
& & $0.03^{+0.03}_{-0.03}$  & $ 0.14^{+ 0.01}_{-0.01}$  & $12^{+15}_{-6}$ &\nodata&\nodata&\nodata&\nodata&$1.13( 68)$ & $ 0.31^{+ 0.02}_{-0.02}$  & $ 0.44^{+ 0.16}_{-0.10}$ \\
\hline
&0505720201 & $0.15^{+0.06}_{-0.06}$ &\nodata&\nodata&\nodata&\nodata & $  2.3^{+  0.3}_{ -0.3}$  & $ 85^{+ 22}_{-17}$ &$1.15( 26)$ & $ 2.6^{+ 0.4}_{-0.4}$  & $ 4.2^{+ 1.0}_{-0.7}$ \\
\cline{2-12}
&0505720301 & $0.00^{+0.05}$ &\nodata&\nodata&\nodata&\nodata & $  1.9^{+  0.3}_{ -0.2}$  & $ 30^{+ 8}_{ -4}$ &$0.96( 14)$ & $ 1.8^{+ 0.3}_{-0.4}$  & $ 1.7^{+ 0.2}_{-0.3}$ \\
\cline{2-12}
&0505720401 & $0.00^{+0.09}$ &\nodata&\nodata&\nodata&\nodata & $  1.3^{+  0.2}_{ -0.2}$  & $ 25^{+  4}_{ -4}$ &$1.32(  8)$ & $ 2.5^{+ 0.7}_{-0.6}$  & $ 2.5^{+ 0.7}_{-0.6}$ \\
\cline{2-12}
&0505720501 & $0.04^{+0.07}_{-0.04}$ &\nodata&\nodata&\nodata&\nodata & $  1.6^{+  0.3}_{ -0.3}$  & $113^{+ 32}_{-24}$ &$0.85( 14)$ & $ 8^{+ 1}_{-1}$  & $ 8^{+ 1}_{-1}$ \\
\cline{2-12}
\multirow{19}{*}{222148} & 0505720601 & $0.13^{+0.06}_{-0.05}$ &\nodata&\nodata&\nodata&\nodata& $ 1.8^{+ 0.2}_{-0.2}$  & $127^{+28}_{-23}$ &$0.76( 37)$ & $ 6.5^{+ 0.9}_{-0.8}$  & $ 8.1^{+ 0.8}_{-0.8}$ \\
\cline{2-12}
&0551690201 & $0.08^{+0.06}_{-0.05}$ &\nodata&\nodata&\nodata&\nodata & $  1.7^{+  0.3}_{ -0.2}$  & $ 73^{+ 19}_{-15}$ &$1.06( 24)$ & $ 4.3^{+ 0.8}_{-0.7}$  & $ 5.0^{+ 0.6}_{-0.6}$ \\
\cline{2-12}
&0551690301 & $0.09^{+0.08}_{-0.06}$ &\nodata&\nodata&\nodata&\nodata & $  1.9^{+  0.4}_{ -0.3}$  & $ 52^{+ 18}_{-13}$ &$0.74( 15)$ & $ 2.5^{+ 0.6}_{-0.5}$  & $ 3.0^{+ 0.5}_{-0.5}$ \\
\cline{2-12}
&0551690401 & $0.1^{+0.1}_{-0.1}$ &\nodata&\nodata&\nodata&\nodata & $  1.7^{+  0.5}_{ -0.5}$  & $ 87^{+ 48}_{-30}$ &\nodata & $ 5^{+ 2}_{-1}$  & $ 6^{+ 2}_{-1}$ \\
\cline{2-12}
&0551690501 & $0.03^{+0.12}_{-0.03}$ &\nodata&\nodata&\nodata&\nodata & $  2.1^{+  0.7}_{ -0.4}$  & $ 47^{+ 24}_{-11}$ &$1.09(  6)$ & $ 2.2^{+ 0.6}_{-0.6}$  & $ 2.5^{+ 1.0}_{-0.5}$ \\
\cline{2-12}
&0551690601 & $0.2^{+0.2}_{-0.2}$ &\nodata&\nodata&\nodata&\nodata & $  2.8^{+  1.1}_{ -0.8}$  & $ 57^{+ 55}_{-25}$ &\nodata & $ 1.2^{+ 0.6}_{-0.4}$  & $ 3^{+ 6}_{-1}$ \\
\cline{2-12}
&0600660201 & $0.10^{+0.06}_{-0.05}$ &\nodata&\nodata&\nodata&\nodata & $  1.8^{+  0.2}_{ -0.2}$  & $ 96^{+ 21}_{-17}$ &$1.06( 27)$ & $ 5.4^{+ 0.9}_{-0.8}$  & $ 6.4^{+ 0.8}_{-0.7}$ \\
\cline{2-12}
&0600660301 & $0.14^{+0.07}_{-0.06}$ &\nodata&\nodata&\nodata&\nodata & $  2.0^{+  0.3}_{ -0.3}$  & $106^{+ 28}_{-22}$ &$1.03( 21)$ & $ 4.3^{+ 0.7}_{-0.7}$  & $ 5.9^{+ 0.9}_{-0.8}$ \\
\cline{2-12}
&0600660401 & $0.10^{+0.07}_{-0.06}$ &\nodata&\nodata&\nodata&\nodata & $  1.8^{+  0.3}_{ -0.2}$  & $ 94^{+ 24}_{-19}$ &$0.42( 21)$ & $ 4.8^{+ 0.8}_{-0.8}$  & $ 5.8^{+ 0.8}_{-0.7}$ \\
\cline{2-12}
&0600660501 & $0.08^{+0.07}_{-0.06}$ &\nodata&\nodata&\nodata&\nodata & $  1.8^{+  0.3}_{ -0.3}$  & $ 90^{+ 28}_{-21}$ &$0.97( 13)$ & $ 4.8^{+ 0.9}_{-0.8}$  & $ 5.7^{+ 0.8}_{-0.7}$ \\
\cline{2-12}
&0600660601 & $0.14^{+0.09}_{-0.08}$ &\nodata&\nodata&\nodata&\nodata & $  2.0^{+  0.4}_{ -0.3}$  & $105^{+ 38}_{-27}$ &$0.90( 14)$ & $ 4.2^{+ 0.9}_{-0.8}$  & $ 5.8^{+ 1.3}_{-0.9}$ \\
\cline{2-12}
&0650560201 & $0.11^{+0.05}_{-0.04}$ &\nodata&\nodata&\nodata&\nodata & $  1.7^{+  0.2}_{ -0.2}$  & $112^{+ 20}_{-17}$ &$1.23( 43)$ & $ 6.3^{+ 0.8}_{-0.8}$  & $ 7.6^{+ 0.7}_{-0.7}$ \\
\cline{2-12}
&0650560301 & $0.10^{+0.04}_{-0.04}$ &\nodata&\nodata&\nodata&\nodata & $  1.8^{+  0.2}_{ -0.1}$  & $113^{+ 18}_{-15}$ &$0.72( 51)$ & $ 5.9^{+ 0.6}_{-0.6}$  & $ 7.2^{+ 0.6}_{-0.6}$ \\
\cline{2-12}
&0650560401 & $0.15^{+0.07}_{-0.06}$ &\nodata&\nodata&\nodata&\nodata & $  2.0^{+  0.3}_{ -0.2}$  & $121^{+ 29}_{-23}$ &$0.88( 27)$ & $ 4.7^{+ 0.8}_{-0.7}$  & $ 6.6^{+ 1.0}_{-0.8}$ \\
\cline{2-12}
&0650560501 & $0.14^{+0.07}_{-0.06}$ &\nodata&\nodata&\nodata&\nodata & $  1.9^{+  0.3}_{ -0.2}$  & $136^{+ 35}_{-27}$ &$1.02( 21)$ & $ 6.3^{+ 1.0}_{-0.9}$  & $ 8.2^{+ 0.9}_{-0.8}$ \\
\cline{2-12}
&0650560601 & $0.08^{+0.04}_{-0.04}$ &\nodata&\nodata&\nodata&\nodata & $  1.8^{+  0.2}_{ -0.2}$  & $114^{+ 19}_{-16}$ &$0.70( 41)$ & $ 6.3^{+ 0.7}_{-0.7}$  & $ 7.4^{+ 0.7}_{-0.7}$ \\
\cline{2-12}
&\multirow{2}{*}{0672130101} & $0.13^{+0.03}_{-0.03}$ &\nodata&\nodata&\nodata&\nodata & $  1.8^{+  0.1}_{ -0.1}$  & $117^{+ 14}_{-12}$ &$0.85( 74)$ & $ 6.2^{+ 0.5}_{-0.4}$  & $ 7.7^{+ 0.4}_{-0.4}$ \\
\cline{3-12}
&& $0.05^{+0.04}_{-0.03}$  & $ 0.6^{+ 0.2}_{-0.1}$  & $0.10^{+0.18}_{-0.07}$  & $1.4^{+1.0}_{-0.3}$  & $0.008^{+0.012}_{-0.007}$ &\nodata&\nodata&$0.85( 71)$ & $ 5.8^{+ 0.8}_{-0.5}$  & $ 6.3^{+ 0.7}_{-0.6}$ \\
\cline{2-12}
&0672130601 & $0.13^{+0.04}_{-0.03}$ &\nodata&\nodata&\nodata&\nodata & $  1.8^{+  0.1}_{ -0.1}$  & $118^{+ 15}_{-13}$ &$0.71( 70)$ & $ 6.2^{+ 0.5}_{-0.5}$  & $ 7.7^{+ 0.4}_{-0.4}$ \\
\cline{2-12}
&0672130701 & $0.11^{+0.03}_{-0.03}$ &\nodata&\nodata&\nodata&\nodata & $  1.7^{+  0.1}_{ -0.1}$  & $111^{+ 14}_{-12}$ &$1.08( 70)$ & $ 6.4^{+ 0.5}_{-0.5}$  & $ 7.7^{+ 0.4}_{-0.4}$ \\
\cline{2-12}
&0672130501 & $0.10^{+0.06}_{-0.05}$ &\nodata&\nodata&\nodata&\nodata & $  1.6^{+  0.2}_{ -0.2}$  & $ 88^{+ 22}_{-17}$ &$1.11( 25)$ & $ 5.6^{+ 0.8}_{-0.8}$  & $ 6.5^{+ 0.6}_{-0.6}$ \\
\hline
222243 & 0405320801 & $0.01^{+0.06}_{-0.01}$  & $ 0.4^{+ 0.1}_{-0.1}$  & $0.3^{+1.1}_{-0.2}$  & $1.5^{+0.1}_{-0.1}$  & $0.019^{+0.005}_{-0.004}$ &\nodata&\nodata&$0.94(118)$ & $ 9.4^{+ 0.7}_{-0.7}$  & $ 9.5^{+ 0.7}_{-0.7}$ \\
\hline
241580 & 0556300101 & $0.2^{+0.1}_{-0.1}$ &\nodata&\nodata&\nodata&\nodata& $ 1.6^{+ 0.2}_{-0.2}$  & $18^{+ 6}_{-4}$ &$1.07( 26)$ & $ 1.1^{+ 0.2}_{-0.2}$  & $ 1.3^{+ 0.2}_{-0.2}$
\enddata 
\tablecomments{Columns: (1) 2XMM-DR3 Unique source index; (2) observation ID; (3) absorption column density; (4) the MCD temperature; (5) the MCD normization $N_{\rm MCD}\equiv ((R_{\rm MCD}/{\rm km})/(D/{\rm 10 kpc}))^2\cos\theta$, where $R_{\rm MCD}$ is the apparent inner disk radius, $D$ is the source distance, $\theta$ is the disk inclination; (6) the BB temperature; (7) the BB normization $N_{\rm  BB}\equiv ((R_{\rm BB}/{\rm km})/(D/{\rm 10 kpc}))^2$, where $R_{\rm BB}$ is the source radius; (8) the PL photon index; (9) the PL normalization; (10)-(11) the 0.3--10 keV absorbed and unabsorbed flux, respectively. All the fits used spectra within 0.3--10 keV. The $N_{\rm H}$ values shared by multiple observations was obtained from our simultaneous fits of these observations with this parameter tied together. Parameters without errors were fixed in the fits.}
\tablenotetext{a}{The $\chi^2$ values are listed only for fits using $\chi^2$ statistic.}
\tablenotetext{b}{These fits included a narrow Gaussian line, with $\sigma_{\rm Ga}$ fixed at 0.01 keV and the best-fitting centroid energy $E_{\rm Ga}=0.87\pm0.02$ keV.}
\tablenotetext{c}{This fit included a narrow Gaussian line, with the best-fitting $E_{\rm Ga}=6.7\pm0.1$ keV, $\sigma_{\rm Ga}=0.2\pm0.1$ keV and $N_{\rm Ga}=(4\pm1)\times 10^{-6}$. The EW is $1.0\pm0.3$ keV. }
\tablenotetext{d}{This fit included a narrow Gaussian line, with the best-fitting $E_{\rm Ga}=6.6\pm0.1$ keV, $\sigma_{\rm Ga}=0.2\pm0.2$ keV and $N_{\rm Ga}=(3\pm1)\times 10^{-6}$. The EW is $0.9\pm0.4$ keV. There is also some residual at energy around 8 keV, which is mostly likely due to the pn instrument background line (Cu-K$_\alpha$). We thus excluded the pn data between 7.8--8.2 keV for this spectral fit. }
\end{deluxetable*}

\tabletypesize{\tiny}
\setlength{\tabcolsep}{0.01in}
\begin{deluxetable*}{rlccccccccccccccccc}
\centering
\tablecaption{The candidate SDSS counterparts \label{tbl:sdssmat}}
\tablewidth{0pt}
\tablehead{SRCID & \colhead{SDSS8} & \colhead{Dxo} & \colhead{cl} &\colhead{$u$} & \colhead{$g$} & \colhead{$r$} & \colhead{$i$} & \colhead{$z$}\\
(1) &\colhead{(2)} &(3) &(4) &(5) &(6) &(7) &(8) & (9)
}
\startdata
101335 &J121028.95+391748.9 &0.2 &6 &22.64$\pm$0.40 &22.22$\pm$0.12 &21.67$\pm$0.10 &21.68$\pm$0.14 &21.07$\pm$0.26 \\ 
101369 &J121034.99+393122.9 &0.6 &3 &22.00$\pm$0.16 &20.49$\pm$0.02 &19.49$\pm$0.01 &19.10$\pm$0.01 &18.79$\pm$0.04 \\ 
107102 &J123103.24+110648.6 &0.5 &3 &22.21$\pm$0.30 &20.75$\pm$0.04 &20.12$\pm$0.03 &19.71$\pm$0.03 &19.45$\pm$0.08 \\ 
174295 &J203353.67+410717.2 &0.1 &6 &25.91$\pm$0.48 &24.47$\pm$0.47 &21.44$\pm$0.07 &19.65$\pm$0.02 &18.28$\pm$0.03 
\enddata 
\tablecomments{Columns are as follows. (1): 2XMM-DR3 Unique source index; (2) SDSS8 source designation; (3) X-ray-optical separation (arcsec) (4) SDSS object class (``3'': galaxy; ``6'': star); (5)-(9): SDSS magnitudes.}
\end{deluxetable*}
\clearpage


\begin{thebibliography}{73}
\expandafter\ifx\csname natexlab\endcsname\relax\def\natexlab#1{#1}\fi

\bibitem[{{Arnaud}(1996)}]{ar1996}
{Arnaud}, K.~A. 1996, in Astronomical Society of the Pacific Conference Series,
  Vol. 101, Astronomical Data Analysis Software and Systems V, ed. G.~H.
  {Jacoby} \& J.~{Barnes}, 17--+

\bibitem[{{Barnard} {et~al.}(2012){Barnard}, {Garcia}, \&
  {Murray}}]{bagamu2012}
{Barnard}, R., {Garcia}, M., \& {Murray}, S.~S. 2012, \apj, 757, 40

\bibitem[{{Barnard} {et~al.}(2008){Barnard}, {Greening}, \&
  {Kolb}}]{bagrko2008}
{Barnard}, R., {Greening}, L.~S., \& {Kolb}, U. 2008, \mnras, 388, 849

\bibitem[{{Bautz} {et~al.}(1998){Bautz}, {Pivovaroff}, {Baganoff}, {Isobe},
  {Jones}, {Kissel}, {Lamarr}, {Manning}, {Prigozhin}, {Ricker}, {Nousek},
  {Grant}, {Nishikida}, {Scholze}, {Thornagel}, \& {Ulm}}]{bapiba1998}
{Bautz}, M.~W., {Pivovaroff}, M., {Baganoff}, F., {et~al.} 1998, in Society of
  Photo-Optical Instrumentation Engineers (SPIE) Conference Series, Vol. 3444,
  Society of Photo-Optical Instrumentation Engineers (SPIE) Conference Series,
  ed. {R.~B.~Hoover \& A.~B.~Walker}, 210--224

\bibitem[{{Bil{\'{\i}}kov{\'a}} {et~al.}(2010){Bil{\'{\i}}kov{\'a}}, {Chu},
  {Gruendl}, \& {Maddox}}]{bichgr2010}
{Bil{\'{\i}}kov{\'a}}, J., {Chu}, Y., {Gruendl}, R.~A., \& {Maddox}, L.~A.
  2010, \aj, 140, 1433

\bibitem[{{Cutri} {et~al.}(2003){Cutri}, {Skrutskie}, {van Dyk}, {Beichman},
  {Carpenter}, {Chester}, {Cambresy}, {Evans}, {Fowler}, {Gizis}, {Howard},
  {Huchra}, {Jarrett}, {Kopan}, {Kirkpatrick}, {Light}, {Marsh}, {McCallon},
  {Schneider}, {Stiening}, {Sykes}, {Weinberg}, {Wheaton}, {Wheelock}, \&
  {Zacarias}}]{cuskva2003}
{Cutri}, R.~M., {Skrutskie}, M.~F., {van Dyk}, S., {et~al.} 2003, VizieR Online
  Data Catalog, 2246, 0

\bibitem[{{de Vaucouleurs} {et~al.}(1991){de Vaucouleurs}, {de Vaucouleurs},
  {Corwin}, {Buta}, {Paturel}, \& {Fouque}}]{dedeco1991}
{de Vaucouleurs}, G., {de Vaucouleurs}, A., {Corwin}, Jr., H.~G., {et~al.}
  1991, {Third Reference Catalogue of Bright Galaxies}, ed. {de Vaucouleurs,
  G., de Vaucouleurs, A., Corwin, H.~G., Jr., Buta, R.~J., Paturel, G., \&
  Fouque, P.}

\bibitem[{{Di Stefano} {et~al.}(2001){Di Stefano}, {Greiner}, {Murray}, \&
  {Garcia}}]{digrmu2001}
{Di Stefano}, R., {Greiner}, J., {Murray}, S., \& {Garcia}, M. 2001, \apjl,
  551, L37

\bibitem[{{Di Stefano} {et~al.}(2010){Di Stefano}, {Kong}, \&
  {Primini}}]{dikopr2010}
{Di Stefano}, R., {Kong}, A., \& {Primini}, F.~A. 2010, New A Rev., 54, 72

\bibitem[{{Done} \& {Gierli{\'n}ski}(2003)}]{dogi2003}
{Done}, C. \& {Gierli{\'n}ski}, M. 2003, \mnras, 342, 1041

\bibitem[{{Endl} {et~al.}(1997){Endl}, {Strassmeier}, \&
  {Kurster}}]{enstku1997}
{Endl}, M., {Strassmeier}, K.~G., \& {Kurster}, M. 1997, \aap, 328, 565

\bibitem[{{Evans} {et~al.}(2010){Evans}, {Primini}, {Glotfelty}, {Anderson},
  {Bonaventura}, {Chen}, {Davis}, {Doe}, {Evans}, {Fabbiano}, {Galle}, {Gibbs},
  {Grier}, {Hain}, {Hall}, {Harbo}, {(Helen He}, {Houck}, {Karovska},
  {Kashyap}, {Lauer}, {McCollough}, {McDowell}, {Miller}, {Mitschang},
  {Morgan}, {Mossman}, {Nichols}, {Nowak}, {Plummer}, {Refsdal}, {Rots},
  {Siemiginowska}, {Sundheim}, {Tibbetts}, {Van Stone}, {Winkelman}, \&
  {Zografou}}]{evprgl2010}
{Evans}, I.~N., {Primini}, F.~A., {Glotfelty}, K.~J., {et~al.} 2010, \apjs,
  189, 37

\bibitem[{{Ezuka} \& {Ishida}(1999)}]{ezis1999}
{Ezuka}, H. \& {Ishida}, M. 1999, \apjs, 120, 277

\bibitem[{{Feng} \& {Soria}(2011)}]{feso2011}
{Feng}, H. \& {Soria}, R. 2011, New A Rev., 55, 166

\bibitem[{{Freedman} {et~al.}(2001){Freedman}, {Madore}, {Gibson}, {Ferrarese},
  {Kelson}, {Sakai}, {Mould}, {Kennicutt}, {Ford}, {Graham}, {Huchra},
  {Hughes}, {Illingworth}, {Macri}, \& {Stetson}}]{frmagi2001}
{Freedman}, W.~L., {Madore}, B.~F., {Gibson}, B.~K., {et~al.} 2001, \apj, 553,
  47

\bibitem[{{Galache} {et~al.}(2007){Galache}, {Garcia}, {Steeghs}, {Torres},
  {Murray}, \& {Williams}}]{gagast2007}
{Galache}, J.~L., {Garcia}, M.~R., {Steeghs}, D., {et~al.} 2007, The
  Astronomer's Telegram, 1171, 1

\bibitem[{{Galleti} {et~al.}(2004){Galleti}, {Federici}, {Bellazzini}, {Fusi
  Pecci}, \& {Macrina}}]{gafebe2004}
{Galleti}, S., {Federici}, L., {Bellazzini}, M., {Fusi Pecci}, F., \&
  {Macrina}, S. 2004, \aap, 416, 917

\bibitem[{{Galloway} {et~al.}(2008){Galloway}, {Morgan}, \&
  {Chakrabarty}}]{gamoch2008}
{Galloway}, D.~K., {Morgan}, E.~H., \& {Chakrabarty}, D. 2008, in American
  Institute of Physics Conference Series, Vol. 1068, American Institute of
  Physics Conference Series, ed. R.~{Wijnands}, D.~{Altamirano}, P.~{Soleri},
  N.~{Degenaar}, N.~{Rea}, P.~{Casella}, A.~{Patruno}, \& M.~{Linares}, 55--62

\bibitem[{{Galloway} {et~al.}(2007){Galloway}, {Morgan}, {Krauss}, {Kaaret}, \&
  {Chakrabarty}}]{gamokr2007}
{Galloway}, D.~K., {Morgan}, E.~H., {Krauss}, M.~I., {Kaaret}, P., \&
  {Chakrabarty}, D. 2007, \apjl, 654, L73

\bibitem[{{Gierli{\'n}ski} \& {Done}(2004)}]{gido2004}
{Gierli{\'n}ski}, M. \& {Done}, C. 2004, \mnras, 349, L7

\bibitem[{{Gierli{\'n}ski} {et~al.}(2008){Gierli{\'n}ski}, {Middleton}, {Ward},
  \& {Done}}]{gimiwa2008}
{Gierli{\'n}ski}, M., {Middleton}, M., {Ward}, M., \& {Done}, C. 2008, \nat,
  455, 369

\bibitem[{{Greiner}(2000)}]{gr2000}
{Greiner}, J. 2000, New Astronomy, 5, 137

\bibitem[{{Hartman} {et~al.}(2009){Hartman}, {Watts}, \&
  {Chakrabarty}}]{hawach2009}
{Hartman}, J.~M., {Watts}, A.~L., \& {Chakrabarty}, D. 2009, \apj, 697, 2102

\bibitem[{{Ho} {et~al.}(2012){Ho}, {Kim}, \& {Terashima}}]{hokite2012}
{Ho}, L.~C., {Kim}, M., \& {Terashima}, Y. 2012, \apjl, 759, L16

\bibitem[{{Hofmann} {et~al.}(2013){Hofmann}, {Pietsch}, {Henze}, {Haberl},
  {Sturm}, {Della Valle}, {Hartmann}, \& {Hatzidimitriou}}]{hopihe2013}
{Hofmann}, F., {Pietsch}, W., {Henze}, M., {et~al.} 2013, \aap, 555, A65

\bibitem[{{Jones} {et~al.}(2009){Jones}, {Read}, {Saunders}, {Colless},
  {Jarrett}, {Parker}, {Fairall}, {Mauch}, {Sadler}, {Watson}, {Burton},
  {Campbell}, {Cass}, {Croom}, {Dawe}, {Fiegert}, {Frankcombe}, {Hartley},
  {Huchra}, {James}, {Kirby}, {Lahav}, {Lucey}, {Mamon}, {Moore}, {Peterson},
  {Prior}, {Proust}, {Russell}, {Safouris}, {Wakamatsu}, {Westra}, \&
  {Williams}}]{joresa2009}
{Jones}, D.~H., {Read}, M.~A., {Saunders}, W., {et~al.} 2009, \mnras, 399, 683

\bibitem[{{Kaplan} {et~al.}(2002){Kaplan}, {van Kerkwijk}, \&
  {Anderson}}]{kavaan2002}
{Kaplan}, D.~L., {van Kerkwijk}, M.~H., \& {Anderson}, J. 2002, \apj, 571, 447

\bibitem[{{King} \& {Pounds}(2003)}]{kipo2003}
{King}, A.~R. \& {Pounds}, K.~A. 2003, \mnras, 345, 657

\bibitem[{{Kong} \& {Di Stefano}(2005)}]{kodi2005}
{Kong}, A.~K.~H. \& {Di Stefano}, R. 2005, \apjl, 632, L107

\bibitem[{{Kong} {et~al.}(2003){Kong}, {Sjouwerman}, {Williams}, {Garcia}, \&
  {Dickel}}]{kosjwi2003}
{Kong}, A.~K.~H., {Sjouwerman}, L.~O., {Williams}, B.~F., {Garcia}, M.~R., \&
  {Dickel}, J.~R. 2003, \apjl, 590, L21

\bibitem[{{Kuerster} \& {Schmitt}(1996)}]{kusc1996}
{Kuerster}, M. \& {Schmitt}, J.~H.~M.~M. 1996, \aap, 311, 211

\bibitem[{{Leahy} {et~al.}(1983){Leahy}, {Darbro}, {Elsner}, {Weisskopf},
  {Kahn}, {Sutherland}, \& {Grindlay}}]{ledael1983}
{Leahy}, D.~A., {Darbro}, W., {Elsner}, R.~F., {et~al.} 1983, \apj, 266, 160

\bibitem[{{Lidskii} \& {Ozernoi}(1979)}]{lioz1979}
{Lidskii}, V.~V. \& {Ozernoi}, L.~M. 1979, Soviet Astronomy Letters, 5, 16

\bibitem[{{Lin} {et~al.}(2011){Lin}, {Carrasco}, {Grupe}, {Webb}, {Barret}, \&
  {Farrell}}]{licagr2011}
{Lin}, D., {Carrasco}, E.~R., {Grupe}, D., {et~al.} 2011, \apj, 738, 52

\bibitem[{{Lin} {et~al.}(2013{\natexlab{a}}){Lin}, {Irwin}, {Godet}, {Webb}, \&
  {Barret}}]{liirgo2013}
{Lin}, D., {Irwin}, J.~A., {Godet}, O., {Webb}, N.~A., \& {Barret}, D.
  2013{\natexlab{a}}, \apjl, 776, L10

\bibitem[{{Lin} {et~al.}(2007){Lin}, {Remillard}, \& {Homan}}]{lireho2007}
{Lin}, D., {Remillard}, R.~A., \& {Homan}, J. 2007, \apj, 667, 1073

\bibitem[{{Lin} {et~al.}(2009){Lin}, {Remillard}, \& {Homan}}]{lireho2009}
---. 2009, \apj, 696, 1257

\bibitem[{{Lin} {et~al.}(2010){Lin}, {Remillard}, \& {Homan}}]{lireho2010}
---. 2010, \apj, 719, 1350

\bibitem[{{Lin} {et~al.}(2012{\natexlab{a}}){Lin}, {Remillard}, {Homan}, \&
  {Barret}}]{lireho2012}
{Lin}, D., {Remillard}, R.~A., {Homan}, J., \& {Barret}, D. 2012{\natexlab{a}},
  \apj, 756, 34

\bibitem[{{Lin} {et~al.}(2012{\natexlab{b}}){Lin}, {Webb}, \&
  {Barret}}]{liweba2012}
{Lin}, D., {Webb}, N.~A., \& {Barret}, D. 2012{\natexlab{b}}, \apj, 756, 27

\bibitem[{{Lin} {et~al.}(2013{\natexlab{b}}){Lin}, {Webb}, \&
  {Barret}}]{liweba2013}
---. 2013{\natexlab{b}}, \apj, 766, 29

\bibitem[{{Liu}(2011)}]{li2011}
{Liu}, J. 2011, \apjs, 192, 10

\bibitem[{{Liu} \& {Bregman}(2005)}]{libr2005}
{Liu}, J.-F. \& {Bregman}, J.~N. 2005, \apjs, 157, 59

\bibitem[{{Marggraf} {et~al.}(2004){Marggraf}, {Bluhm}, \& {de
  Boer}}]{mablde2004}
{Marggraf}, O., {Bluhm}, H., \& {de Boer}, K.~S. 2004, \aap, 416, 251

\bibitem[{{Massey} {et~al.}(2007){Massey}, {Olsen}, {Hodge}, {Jacoby},
  {McNeill}, {Smith}, \& {Strong}}]{maolho2007}
{Massey}, P., {Olsen}, K.~A.~G., {Hodge}, P.~W., {et~al.} 2007, \aj, 133, 2393

\bibitem[{{McClintock} \& {Remillard}(2006)}]{mcre2006}
{McClintock}, J.~E. \& {Remillard}, R.~A. 2006, {Compact Stellar X-ray Sources,
  ed. W. Lewin and M. van der Klis (Cambridge: Cambridge Univ. Press)},
  157--213

\bibitem[{{Mereghetti} {et~al.}(2010){Mereghetti}, {Krachmalnicoff}, {La
  Palombara}, {Tiengo}, {Rauch}, {Haberl}, {Filipovi{\'c}}, \&
  {Sturm}}]{mekr2010}
{Mereghetti}, S., {Krachmalnicoff}, N., {La Palombara}, N., {et~al.} 2010,
  \aap, 519, A42

\bibitem[{{Miller} {et~al.}(2004){Miller}, {Fabian}, \& {Miller}}]{mifami2004}
{Miller}, J.~M., {Fabian}, A.~C., \& {Miller}, M.~C. 2004, \apjl, 614, L117

\bibitem[{{Miniutti} {et~al.}(2013){Miniutti}, {Saxton},
  {Rodr{\'{\i}}guez-Pascual}, {Read}, {Esquej}, {Colless}, {Dobbie}, \&
  {Spolaor}}]{misaro2013}
{Miniutti}, G., {Saxton}, R.~D., {Rodr{\'{\i}}guez-Pascual}, P.~M., {et~al.}
  2013, \mnras, 433, 1764

\bibitem[{{Mitsuda} {et~al.}(1984){Mitsuda}, {Inoue}, {Koyama}, {Makishima},
  {Matsuoka}, {Ogawara}, {Suzuki}, {Tanaka}, {Shibazaki}, \&
  {Hirano}}]{miinko1984}
{Mitsuda}, K., {Inoue}, H., {Koyama}, K., {et~al.} 1984, \pasj, 36, 741

\bibitem[{{Monet} {et~al.}(2003){Monet}, {Levine}, {Canzian}, {Ables}, {Bird},
  {Dahn}, {Guetter}, {Harris}, {Henden}, {Leggett}, {Levison}, {Luginbuhl},
  {Martini}, {Monet}, {Munn}, {Pier}, {Rhodes}, {Riepe}, {Sell}, {Stone},
  {Vrba}, {Walker}, {Westerhout}, {Brucato}, {Reid}, {Schoening}, {Hartley},
  {Read}, \& {Tritton}}]{moleca2003}
{Monet}, D.~G., {Levine}, S.~E., {Canzian}, B., {et~al.} 2003, \aj, 125, 984

\bibitem[{{Mukai} {et~al.}(2003){Mukai}, {Pence}, {Snowden}, \&
  {Kuntz}}]{mupesn2003}
{Mukai}, K., {Pence}, W.~D., {Snowden}, S.~L., \& {Kuntz}, K.~D. 2003, \apj,
  582, 184

\bibitem[{{Muno} {et~al.}(2004){Muno}, {Arabadjis}, {Baganoff}, {Bautz},
  {Brandt}, {Broos}, {Feigelson}, {Garmire}, {Morris}, \&
  {Ricker}}]{muarba2004}
{Muno}, M.~P., {Arabadjis}, J.~S., {Baganoff}, F.~K., {et~al.} 2004, \apj, 613,
  1179

\bibitem[{{Muno} {et~al.}(2005){Muno}, {Pfahl}, {Baganoff}, {Brandt}, {Ghez},
  {Lu}, \& {Morris}}]{mupfba2005}
{Muno}, M.~P., {Pfahl}, E., {Baganoff}, F.~K., {et~al.} 2005, \apjl, 622, L113

\bibitem[{{Ohsuga} {et~al.}(2005){Ohsuga}, {Mori}, {Nakamoto}, \&
  {Mineshige}}]{ohmona2005}
{Ohsuga}, K., {Mori}, M., {Nakamoto}, T., \& {Mineshige}, S. 2005, \apj, 628,
  368

\bibitem[{{Pedlar} {et~al.}(1992){Pedlar}, {Howley}, {Axon}, \&
  {Unger}}]{pehoax1992}
{Pedlar}, A., {Howley}, P., {Axon}, D.~J., \& {Unger}, S.~W. 1992, \mnras, 259,
  369

\bibitem[{{Poutanen} {et~al.}(2007){Poutanen}, {Lipunova}, {Fabrika},
  {Butkevich}, \& {Abolmasov}}]{polifa2007}
{Poutanen}, J., {Lipunova}, G., {Fabrika}, S., {Butkevich}, A.~G., \&
  {Abolmasov}, P. 2007, \mnras, 377, 1187

\bibitem[{{Pravdo} {et~al.}(2009){Pravdo}, {Tsuboi}, {Suzuki}, {Thompson}, \&
  {Rebull}}]{prtssu2009}
{Pravdo}, S.~H., {Tsuboi}, Y., {Suzuki}, Y., {Thompson}, T.~J., \& {Rebull}, L.
  2009, \apj, 690, 850

\bibitem[{{Ramsay} {et~al.}(2004){Ramsay}, {Cropper}, {Mason}, {C{\'o}rdova},
  \& {Priedhorsky}}]{racrma2004}
{Ramsay}, G., {Cropper}, M., {Mason}, K.~O., {C{\'o}rdova}, F.~A., \&
  {Priedhorsky}, W. 2004, \mnras, 347, 95

\bibitem[{{Rees}(1988)}]{re1988}
{Rees}, M.~J. 1988, \nat, 333, 523

\bibitem[{{Reis} {et~al.}(2012){Reis}, {Miller}, {Reynolds}, {G{\"u}ltekin},
  {Maitra}, {King}, \& {Strohmayer}}]{remire2012}
{Reis}, R.~C., {Miller}, J.~M., {Reynolds}, M.~T., {et~al.} 2012, Science, 337,
  949

\bibitem[{{Remillard} \& {McClintock}(2006)}]{remc2006}
{Remillard}, R.~A. \& {McClintock}, J.~E. 2006, \araa, 44, 49

\bibitem[{{Sakano} {et~al.}(2005){Sakano}, {Warwick}, {Decourchelle}, \&
  {Wang}}]{sawade2005}
{Sakano}, M., {Warwick}, R.~S., {Decourchelle}, A., \& {Wang}, Q.~D. 2005,
  \mnras, 357, 1211

\bibitem[{{Samus} {et~al.}(2009){Samus}, {Durlevich}, \& {et al.}}]{sadu2009}
{Samus}, N.~N., {Durlevich}, O.~V., \& {et al.} 2009, VizieR Online Data
  Catalog, 1, 2025

\bibitem[{{Sana} {et~al.}(2006){Sana}, {Gosset}, {Rauw}, {Sung}, \&
  {Vreux}}]{sagora2006}
{Sana}, H., {Gosset}, E., {Rauw}, G., {Sung}, H., \& {Vreux}, J.-M. 2006, \aap,
  454, 1047

\bibitem[{{Soleri} {et~al.}(2010){Soleri}, {Fender}, {Tudose}, {Maitra},
  {Bell}, {Linares}, {Altamirano}, {Wijnands}, {Belloni}, {Casella},
  {Miller-Jones}, {Muxlow}, {Klein-Wolt}, {Garrett}, \& {van der
  Klis}}]{sofetu2010}
{Soleri}, P., {Fender}, R., {Tudose}, V., {et~al.} 2010, \mnras, 406, 1471

\bibitem[{{Stanek} \& {Garnavich}(1998)}]{stga1998}
{Stanek}, K.~Z. \& {Garnavich}, P.~M. 1998, \apjl, 503, L131

\bibitem[{{Sugizaki} {et~al.}(2001){Sugizaki}, {Mitsuda}, {Kaneda},
  {Matsuzaki}, {Yamauchi}, \& {Koyama}}]{sumika2001}
{Sugizaki}, M., {Mitsuda}, K., {Kaneda}, H., {et~al.} 2001, \apjs, 134, 77

\bibitem[{{Terashima} {et~al.}(2012){Terashima}, {Kamizasa}, {Awaki}, {Kubota},
  \& {Ueda}}]{tekaaw2012}
{Terashima}, Y., {Kamizasa}, N., {Awaki}, H., {Kubota}, A., \& {Ueda}, Y. 2012,
  \apj, 752, 154

\bibitem[{{Watson} {et~al.}(2009){Watson}, {Schr{\"o}der}, {Fyfe}, {Page},
  {Lamer}, {Mateos}, {Pye}, {Sakano}, {Rosen}, {Ballet}, {Barcons}, {Barret},
  {Boller}, {Brunner}, {Brusa}, {Caccianiga}, {Carrera}, {Ceballos}, {Della
  Ceca}, {Denby}, {Denkinson}, {Dupuy}, {Farrell}, {Fraschetti}, {Freyberg},
  {Guillout}, {Hambaryan}, {Maccacaro}, {Mathiesen}, {McMahon}, {Michel},
  {Motch}, {Osborne}, {Page}, {Pakull}, {Pietsch}, {Saxton}, {Schwope},
  {Severgnini}, {Simpson}, {Sironi}, {Stewart}, {Stewart}, {Stobbart}, {Tedds},
  {Warwick}, {Webb}, {West}, {Worrall}, \& {Yuan}}]{wascfy2009}
{Watson}, M.~G., {Schr{\"o}der}, A.~C., {Fyfe}, D., {et~al.} 2009, \aap, 493,
  339

\bibitem[{{Webb} \& {Barret}(2007)}]{weba2007}
{Webb}, N.~A. \& {Barret}, D. 2007, \apj, 671, 727

\bibitem[{{White} {et~al.}(1994){White}, {Giommi}, \& {Angelini}}]{whgian1994}
{White}, N.~E., {Giommi}, P., \& {Angelini}, L. 1994, \iaucirc, 6100, 1

\bibitem[{{Wijnands} {et~al.}(2006){Wijnands}, {in't Zand}, {Rupen},
  {Maccarone}, {Homan}, {Cornelisse}, {Fender}, {Grindlay}, {van der Klis},
  {Kuulkers}, {Markwardt}, {Miller-Jones}, \& {Wang}}]{wiinru2006}
{Wijnands}, R., {in't Zand}, J.~J.~M., {Rupen}, M., {et~al.} 2006, \aap, 449,
  1117

\end{thebibliography}
\end{document}